\documentclass[final]{siamart190516}

\usepackage{amsfonts,amsmath,amsxtra,amssymb}
\usepackage{wrapfig,graphics,graphicx,xcolor,epstopdf}
\usepackage{tabularx, multirow,makecell, diagbox,fancybox, lscape,listings,float,rotating,url}
\usepackage{booktabs}
\usepackage[titletoc,page]{appendix}
\usepackage{tcolorbox}
\usepackage{siunitx} 
\allowdisplaybreaks[1]
\newsiamremark{remark}{Remark}

\usepackage[caption=false]{subfig}

\def \W {\textit{Wolbachia~}}
\def \Wns {\textit{Wolbachia}}

\newcommand{\R}{\mathbb{R}}
\newcommand{\dt}{\Delta t}

\usepackage[norefs,nocites]{refcheck}
\makeatletter
\newcommand{\refcheckize}[1]{%
  \expandafter\let\csname @@\string#1\endcsname#1%
  \expandafter\DeclareRobustCommand\csname relax\string#1\endcsname[1]{%
    \csname @@\string#1\endcsname{##1}\@for\@temp:=##1\do{\wrtusdrf{\@temp}\wrtusdrf{{\@temp}}}}%
  \expandafter\let\expandafter#1\csname relax\string#1\endcsname
}
\newcommand{\refcheckizetwo}[1]{%
  \expandafter\let\csname @@\string#1\endcsname#1%
  \expandafter\DeclareRobustCommand\csname relax\string#1\endcsname[2]{%
    \csname @@\string#1\endcsname{##1}{##2}\wrtusdrf{##1}\wrtusdrf{{##1}}\wrtusdrf{##2}\wrtusdrf{{##2}}}%
  \expandafter\let\expandafter#1\csname relax\string#1\endcsname
}
\makeatother

\refcheckize{\cref}
\refcheckize{\Cref}
\usepackage{enumitem} 

\newcommand{\TheTitle}{Modeling spatial waves of Wolbachia invasion for controlling mosquito-borne diseases} 
\newcommand{\TheAuthors}{Zhuolin Qu, Tong Wu, and James M. Hyman}

\headers{Modeling spatial waves of Wolbachia invasion}{\TheAuthors}

\title{{\TheTitle}\thanks{Submitted to the editors DATE.}}
\author{Zhuolin Qu\thanks{Department of Mathematics, University of Texas at San Antonio (\email{zhuolin.qu@utsa.edu},\email{tong.wu@utsa.edu})}
  \and Tong Wu\footnotemark[2] \and James M. Hyman\thanks{Department of Mathematics, Tulane University (\email{mhyman@tulane.edu})}}
  
\newif\ifnotesw \noteswtrue

\makeatletter 
\newcommand*{\l@section}{}
\newcommand*{\l@subsection}{}
\newcommand*{\l@subsubsection}{}
\newcommand*{\l@paragraph}{}
\newcommand*{\l@subparagraph}{}
\newcommand*{\l@figure}{}
\newcommand*{\l@table}{}
\makeatother                               
\usepackage[subfigure]{tocloft}
\makeatletter
\renewcommand{\contentsname}{Contents}
\renewcommand{\tableofcontents}{%
  \section*{\contentsname}
  \@starttoc{toc}
}
\renewcommand{\thepage}{\arabic{page}}
\makeatother

\begin{document}
\maketitle  
\begin{abstract}
\W is a natural bacterium that can infect mosquitoes and reduce their ability to transmit mosquito-borne diseases, such as dengue fever, Zika, and chikungunya. Field trials and modeling studies have shown that the fraction of infection among the mosquitoes must exceed a threshold level for the infection to persist. To capture this threshold, it is critical to consider the spatial heterogeneity in the distributions of the infected and uninfected mosquitoes, which is created by the local release of the infected mosquitoes. We develop and analyze partial differential equation (PDE) models to study the invasion dynamics of \W infection among mosquitoes in the field. Our reaction-diffusion-type models account for both the complex vertical transmission and the spatial mosquito dispersion. We characterize the threshold for a successful invasion, which is a bubble-shaped profile, called the ``critical bubble''. The critical bubble is optimal in its release size compared to other spatial profiles in a one-dimensional landscape.  The fraction of infection near the release center is higher than the threshold level for the corresponding homogeneously mixing ODE models.  We show that the proposed spatial models give rise to the traveling waves of \W infection when above the threshold. We quantify how the threshold condition and traveling-wave velocity depend on the diffusion coefficients and other model parameters. Numerical studies for different scenarios are presented to inform the design of release strategies.
\end{abstract} 
 
\begin{keywords}
mosquito-borne diseases, Wolbachia, invasion, threshold condition, traveling wave, model reduction
\end{keywords}

\begin{AMS}
93A30, 35K57, 35C07, 92D30
\end{AMS}


\section{Introduction}
\W is a rising mitigation strategy to control the spread of mosquito-borne diseases, such as dengue fever, Zika, and chikungunya. The primary vector for transmitting these viral diseases is the \textit{Aedes aegypti} mosquito, and the \Wns-infected \textit{Aedes aegypti} mosquitoes are less capable of spreading these diseases \cite{bian2010endosymbiotic,dutra2016wolbachia,van2012impact}. Ongoing field trials have demonstrated a significant reduction in dengue incidence after releasing the infected mosquitoes. In the past five years, this approach has resulted in the near-elimination of local dengue cases in Cairns and Townsville, Australia \cite{ryan2020establishment}. Recently, in Yogyakarta City, Indonesia, there was a 76\% reduction in the dengue incidence was announced after \W deployment \cite{indriani2020reduced}. Similar city-wide trials are being carried out in Rio de Janeiro in Brazil and Bello and Medell{\'i}n in Colombia.

It is challenging to sustain \Wns-infection in the wild \textit{Aedes} mosquitoes. \W infection leads to the fitness cost among the female mosquitoes, and the infection may also be limited by the maternal transmission efficiency. Population cage experiments of mixing mosquitoes demonstrated that there exists a minimal infection threshold to have a persisting \W infection in the mosquito population \cite{axford2016fitness}. 

Homogeneous mixing ordinary differential equation (ODE) models of different scales have been developed to quantify the threshold conditions for \W invasion. In \cite{koiller2014aedes}, a detailed compartmental model of 13 ODEs was proposed that includes the egg, larvae, and pupae stage of the immature mosquitoes. The threshold condition is analyzed as a backward bifurcation with an unstable coexistence equilibrium of infected and uninfected groups. In \cite{qu2018modeling}, a 9-ODE model was developed that includes combined aquatic stages, and the threshold condition was analyzed for the perfect and imperfect maternal transmission rate. Hughes et al \cite{hughes2013modelling} derived a host-vector-\W model to quantify the threshold condition for different strains of \W (wAlbB, wMel, and wMelPop) in eliminating dengue transmission. In \cite{xue2018comparing}, a host-vector model was developed to compare the effectiveness of wAlbB and wMel strains of \W to control the spread of dengue, Zika, and chikungunya viruses after it is established in the field.

Most \W models ignore the role that heterogeneous spatial distributions of the infected mosquitoes can have in establishing a stable infection. The threshold estimates by the ODE models are for an ideally controlled situation where infected and uninfected mosquitoes are homogeneously mixed. Even in the absence of any environmental variation, the wind and flight pattern of the released infected mosquitoes can cause spatial variations in the fraction of infection. When infected mosquitoes are released in the wild, although the local infection level may exceed the threshold near the release site, it can be below the threshold and not sustainable near the edges. Field trials have reported the collapse of infection due to the immigration of natural mosquitoes from nearby regions \cite{schmidt2017Local,jiggins2017spread}. Extending the ODE models to PDE models can account for the heterogeneous spatial dynamics, which can help design the field trials and better predict the faith of the field release due to the threshold effect.

Due to the difficulty of analyzing complex high-dimensional PDEs, most previous spatial models were derived based on heuristics and with strong assumptions to produce physically realistic solutions.
In \cite{barton2011spatial}, a reaction-diffusion type spatial model was proposed that considers \Wns-induced cytoplasmic incompatibility and fitness cost. They used a cubic approximation for the vertical transmission of \W and observed traveling wave solutions in this simple heuristic model. The idea of a threshold introduction size for wave initiation was illustrated and derived for the approximated equation. In \cite{lewis1993waves}, a two-equation spatial model was proposed for an alternative biological control, sterile insect technique, where sterilized insects are released to create an extinction wave. A one-equation model was analyzed for its traveling wave solution, assuming that the sterile population is maintained at a constant density in space.

Qu et al. \cite{qu2019generating} derived a hierarchy of reduced systems of 7, 4, and 2 ODEs from a more detailed 9-ODE model \cite{qu2018modeling} with different resolutions. The reduced models captured the biologically relevant effects, such as the basic reproductive number, bifurcation dynamics, and threshold condition for the more complex model. By starting with a detailed model where all of the parameters have biological relevance, the parameters in the reduced models can be expressed in terms of these original meaningful parameters. In this paper, we derive spatial models for \W invasion with spatial dynamics based on this reduced 2-ODE model. 

As a preliminary investigation, we consider the un-directional mosquitoes dispersion only through the diffusion approximation \cite{takahashi2005Mathematical}. The resulting reaction-diffusion type spatial models account for both the complex vertical transmission dynamic that is inherited from the 2-ODE model (reaction term) and horizontal spatial diffusion. We will identify the threshold conditions for a successful \W invasion given a local release of infection in this simplified PDE model. Specifically, we define the threshold in terms of a natural balanced state between the local reproduction growth (reaction) and mosquito dispersion (diffusion), referred to as ``critical bubble'', and we will compare it with what's been identified in the spatially homogeneous setting. Additionally, when the fraction of infection is above the threshold condition, the proposed spatial models give rise to a traveling wave of \W infection that invades into the zero-infection region with a constant velocity.

After briefly reviewing the 2-ODE model that we based on (\cref{sec:2-ODE}), we propose the extended 2-PDE model (\cref{sec:2-PDE}), and we describe how it can be reduced to a 1-PDE model that maintains the bistable behavior (\cref{sec:1-PDE}). This scalar PDE model is much easier to analyze and provides insight into understanding the dynamics of the 2-PDE. We study the threshold condition (\cref{sec:threshold}) and the traveling wave solution (\cref{sec:travel}) for both spatial models and compare them against each other. We then consider the practical aspects of how these models could be used to inform the design of the field release strategies (\cref{sec:323,sec:design}), as well as the sensitivity analysis on the model parameters (\cref{sec:SA}).

\section{\W Transmission Models}
We base our spatial models on a 2-ODE model that is derived from a detailed 9-ODE model \cite{qu2019generating}. The complex nonlinear growth terms 2-ODE model retains the \W maternal transmission dynamics of the original 9-ODE model. The 2-PDE model extends these dynamics to include one-dimensional diffusion of the mosquitoes. This simple extension generates nontrivial wave invasion dynamics and significantly complicates the derivation and understanding of the threshold conditions for establishing a sustainable \W infection.

\subsection{Review of 2-ODE model\label{sec:2-ODE}}
We start with a 2-ODE model \cite{qu2019generating} for \Wns-free female mosquitoes, $F^u(t)$, and \Wns-infected female mosquitoes, $F^w(t)$,
\begin{equation}
\begin{aligned}
\frac{d F^{u}}{d t} &= b_f \phi_u''\,\frac{F^u}{F^u+\frac{\mu_{fw}'}{\mu_{fu}'}F^w} \left(1-\frac{F^u+F^w}{K_f}\right) F^u\\
&\qquad\qquad\qquad \qquad\quad+v_u b_f \phi_w'' \left(1-\frac{F^u+F^w}{K_f}\right) F^w  -\mu_{fu}' F^u\,,\\
\frac{d  F^{w}}{d t} &= v_w b_f \phi_w'' \left(1-\frac{F^u+F^w}{K_f}\right) F^w-\mu_{fw}' F^w\,\label{eq:ODE2}.
\end{aligned}
\end{equation}
The parameters are defined in terms of the biologically relevant parameters from the original 9-ODE model (see \cref{tab:parameter_all}). We have retained the notations from the original paper for readers' convenience. 

\begin{table}[htbp]
\centering
\caption{Model parameters and their baseline values. All the parameters in the 2-ODE model are defined in terms of the biologically relevant parameters from the original 9-ODE model \cite{qu2019generating}. All the rates have dimension $day^{-1}$.}
\label{tab:parameter_all}
\begin{tabular}{@{}llll@{}}
\toprule
\multicolumn{2}{l}{Biological relevant parameters (9-ODE)} & Baseline & Reference \\ \midrule
$b_f$ & Female birth probability & 0.5&\cite{tun2000effects}\\
$v_w$ &Maternal transmission rate& 0.95& \cite{walker2011wmel}\\
$v_u$ &$=1-v_w$ & 0.05&\cite{walker2011wmel}\\
$\sigma$ &Per capita mating rate & 1 &  \cite{schoof1967mating}\\
$\phi_u$ &Per capita egg-laying rate for $F_{pu}$  & 13 & \cite{hoffmann2014stability,mcmeniman2009stable,mcmeniman2010virulent}\\
$\phi_w$ &Per capita egg-laying rate for $F_{pw}$  & 11 & \cite{hoffmann2014stability,walker2011wmel}\\
$\psi$ &Per capita development rate & 1/8.75  & \cite{hoffmann2014stability,walker2011wmel}\\
$\mu_a$& Death rate for $A_u$ or $A_w$ & 0.02 & \cite{hoffmann2014stability,mcmeniman2010virulent,walker2011wmel}\\
$\mu_{fu}$ &Death rate for $F_u$ and $F_{pu}$ & 1/17.5 & \cite{mcmeniman2009stable,styer2007mortality}\\
$\mu_{fw}$ &Death rate for $F_w$ and $F_{pw}$ & 1/15.8 &  \cite{walker2011wmel} \\
$K_a$ &Carrying capacity of aquatic stage & $2\times 10^5$ & Assumed\\ \midrule
\multicolumn{2}{l}{Reduced parameters (2-ODE)} & Baseline & Definition \cite{qu2019generating}  \\ \midrule
$\phi_u''$ & Per capita reproduction rate for $F^u$ & $7.0$ &$ \frac{\psi}{\psi+\mu_a} \frac{\psi}{\psi+\mu_{fu}} \frac{\sigma}{\sigma+\mu_{fu}} \phi_u$ \\
$\phi_w''$ & Per capita reproduction rate for $F^w$ & $5.7$ &$v_w \frac{\psi}{\psi+\mu_a} \frac{\psi}{\psi+\mu_{fw}} \frac{\sigma}{\sigma+\mu_{fw}} \phi_w$ \\
$\mu_{fu}'$ & Death rate for $F^u$& $1/26.25$ &$\frac{\psi}{\psi+\mu_{fu}} \mu_{fu}$\\
$\mu_{fw}'$ & Death rate for $F^w$ & $1/24.55$ &$\frac{\psi}{\psi+\mu_{fw}}\mu_{fw}$\\
$K_f$ &Carrying capacity for females& $3\!\times \!10^5$ &$b_f \Big(1+\frac{\psi}{\mu_{fu}}\Big) K_a$\\
$D_1$ & Diffusion coefficient for $F^u$ ($m^2$/day) & $1.25\!\times\!10^4$ & \cite{silva2020modeling,takahashi2005Mathematical}\\
$D_2$ & Diffusion coefficient for $F^w$ ($m^2$/day) & $1.25\!\times\! 10^4$ & \cite{silva2020modeling,takahashi2005Mathematical}\\
\bottomrule
\end{tabular}
\end{table}

The 2-ODE model \cref{eq:ODE2} describes the complex maternal transmission of \W infection: for \Wns-infected females, $F^w$, a fraction, $v_w$, of their offspring, $\phi_w''$, are infected. About $b_f \approx 1/2$ of the offspring are then developed into the new generation of infected females. This process corresponds to the first nonlinear birth term in the $F^w$ equation. During the maternal transmission, leakage may happen, with probability $v_u = 1-v_w$. This leads to the production of uninfected female mosquitoes (the second nonlinear birth term in the $F^u$ equation). 

Only the uninfected female mosquitoes, $F^u$, who mate with the uninfected males, with probability $m_u = F^u/(F^u+\frac{\mu_{fw}'}{\mu_{fu}'}F^w)$, can produce uninfected offspring (the first nonlinear birth term in $F^u$ equation). When they mate with infected males, with probability $1-m_u$, no viable offspring will be reproduced due to the cytoplasmic incompatibility caused by the \Wns-infection. All the birth terms are regularized by the carrying capacity, $K_f$. \W infection may also impose fitness cost to the female life traits, such as shorter lifespan (or a larger death rate, $\mu_{fw}'>\mu_{fu}'$) and reduced reproduction rate ($\phi_w''<\phi_u''$). 

The 2-ODE model preserves the key biological quantities related to the \W invasion dynamics, such as the basic reproductive number $\R_0$ and threshold condition for a sustained \W infection. Qu and Hyman \cite{qu2019generating} provided a detailed description of the reduction process and the comparison among different reduced models, and we summarize the key findings below. For simplicity of the presentation, we first present the case of perfect maternal transmission rate, $v_w=1$, in the main text. This is also a desired property for field release, where strains (such as wMel) with less fit-cost and high maternal transmission rate can better facilitate the process. We will discuss the imperfect maternal transmission in \cref{sec:imperfect} and the main conclusions are summarized in \cref{sec:appendix}. 

\begin{figure}[htbp]
\centering
\includegraphics[width=0.5\textwidth]{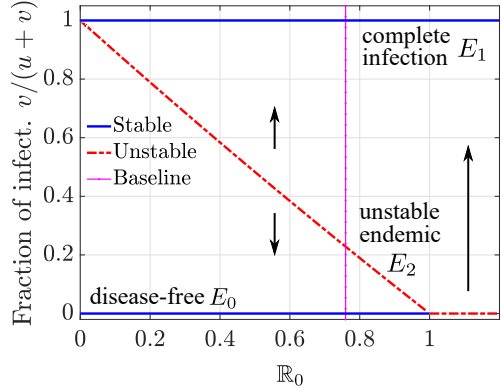}
\caption{Bifurcation plot for the 2-ODE model given a perfect maternal transmission. The unstable steady state $E_2$ corresponds to the threshold condition for a successful \W invasion into a homogeneously mixing mosquito population. At the baseline (\cref{tab:parameter_all} except $v_w=1$), the basic reproductive number $\R_0=0.76<1$, and threshold infection level = $22.84\%$.  \label{fig:bifur}}
\end{figure}

Given the perfect maternal transmission, the basic reproductive number for the ODE model \cref{eq:ODE2} is given by $\R_0 = (\mu_{fu}'\phi_w'')/(\mu_{fw}'\phi_u'')\,.$
Although near the baseline scenario (\cref{tab:parameter_all}), $\R_0<1$, which suggests that introducing a small \W infection might be eliminated due to the fitness cost, the ODE system has a backward bifurcation that identifies a critical threshold condition for a successful invasion. As shown in \cref{fig:bifur}, the system has a stable \Wns-free steady state, $E_0$, a stable \Wns-endemic steady state, $E_1$, and an unstable lower-infection steady state $E_2$, where uninfected and infected mosquitoes coexist. The $E_2$ steady state serves as the bifurcating point or the threshold infection level, above which the infection takes off and approaches the endemic steady state $E_1$ and below which the infection dies out.

\subsection{The 2-PDE model\label{sec:2-PDE}}
\textit{Aedes aegypti} mosquitoes, especially the adult females, make local flights in search of food or places for oviposition. This random and unidirectional movement could be approximated by a diffusion process \cite{takahashi2005Mathematical}. We extend the 2-ODE model \cref{eq:ODE2} to a 2-PDE spatial model, and we define the diffusion coefficients $D_1$ and $D_2$ for the uninfected and infected mosquitoes, respectively, which measure the mean squared displacement of the mosquito flights per day. The extended spatial model, under the perfect maternal transmission, is:
\begin{equation}
\begin{aligned}
\frac{\partial F^{u}}{\partial t} &= b_f \phi_u''\,\frac{F^u}{F^u+\frac{\mu_{fw}'}{\mu_{fu}'}F^w} \left(1-\frac{F^u+F^w}{K_f}\right) F^u -\mu_{fu}' F^u+\nabla\cdot (D_1 \nabla  F^u)\,,\\ 
\frac{\partial  F^{w}}{\partial t} &= b_f \phi_w'' \left(1-\frac{F^u+F^w}{K_f}\right) F^w-\mu_{fw}' F^w+\nabla\cdot (D_2 \nabla  F^w)\,\label{eq:ODE2p},
\end{aligned}
\end{equation}
where $F^u(x,t)$ and $F^w(x,t)$ are population sizes for the uninfected and infected female mosquitoes. The diffusion coefficients $D_1(x)$ and $D_2(x)$ may be location-dependent to reflect the spatial heterogeneity in the environment. For this study, we focus on the basic case where these coefficients are constants. To simplify the presentation of the analysis, we nondimensionalize the system \cref{eq:ODE2p} and introduce the new coefficients and state variables as follows,
\begin{equation}\label{eq:trans}
\begin{aligned}
&u = \frac{F^u}{K_f},\quad v= \frac{F^w}{K_f}, \quad t^\ast=t\,b_f\phi_u'',\quad x^\ast=x\left(\frac{b_f\phi_u''}{D_1}\right)^{1/2}\,,\\
&a = \frac{\phi_w''}{\phi_u''}<1,\quad  b=\frac{\mu_{fu}'}{b_f\phi_u''}<1,\quad d = \frac{\mu_{fw}'}{\mu_{fu}'}>1,\quad D = \frac{D_2}{D_1}\,.
\end{aligned}
\end{equation}
Dropping the asterisks for notational simplicity, we rewrite the \cref{eq:ODE2p} as
\begin{equation}\label{eq:2PDE}
\begin{aligned}
u_t &= \frac{u}{u+d\,v} (1-u-v)u  -b\,u+ u_{xx}\,,\\
v_t &= a (1-u-v) v-b\,d\,v+Dv_{xx}\,,
\end{aligned}
\end{equation}
subject to initial condition $u(x,0)=\Phi_u(x),~~ v(x,0)=\Phi_v(x).$ As derived in the ODE case \cite{qu2019generating} (see also \cref{fig:bifur}), there are three spatially homogeneous steady states:
\begin{equation*}
E_0 = (u_0,0) = (1-b,0)\,,\quad E_1 = (0,v_1) =  \Big(0,1-\displaystyle\frac{b\,d}{a}\Big)\,, \quad E_2 = (u_2,v_2)\,,
\end{equation*}
where $ u_2 = \frac{a\,d}{a\,d+d-a}\big(1-\frac{b\,d}{a}\big)$, and $v_2=\frac{d-a}{a\,d+d-a}\big(1-\frac{b\,d}{a}\big)$.

\subsection{Two-stage invasion dynamics for spatial models}
For the spatial model \cref{eq:2PDE}, we are interested in identifying a threshold condition for the \Wns-infected mosquitoes to invade into a local region. When the fraction of infection is above this threshold, the invasion is sustained and the infection wave propagates across the field. We consider biologically relevant release covering a bounded region with a compact support.

When \Wns-infected mosquitoes are introduced to an empty field, where no mosquitoes are present ($u=0$), the system \cref{eq:2PDE} reduces to 
\begin{equation*}
v_t = a(1-v)v -bd\,v+ D v_{xx}=a\left(1-\frac{b\,d}{a}-v\right)v + D v_{xx}\,.
\end{equation*}
This PDE is equivalent to the well-known Fisher's equation. The roots of the quadratic birth term give two spatially uniform steady states of the equation: extinction of mosquitoes and maximum sustainable mosquitoes.  Kolmogorov et al. \cite{kolmogorov1937etude} showed that given a compact initial condition, the invasion wave happens, and the solution of Fisher's equation converges to a traveling wave solution, sweeping across the domain with a fixed wave speed and joining the two steady states. 

\begin{figure}[ht!]
\centering
\includegraphics[width=0.49\textwidth]{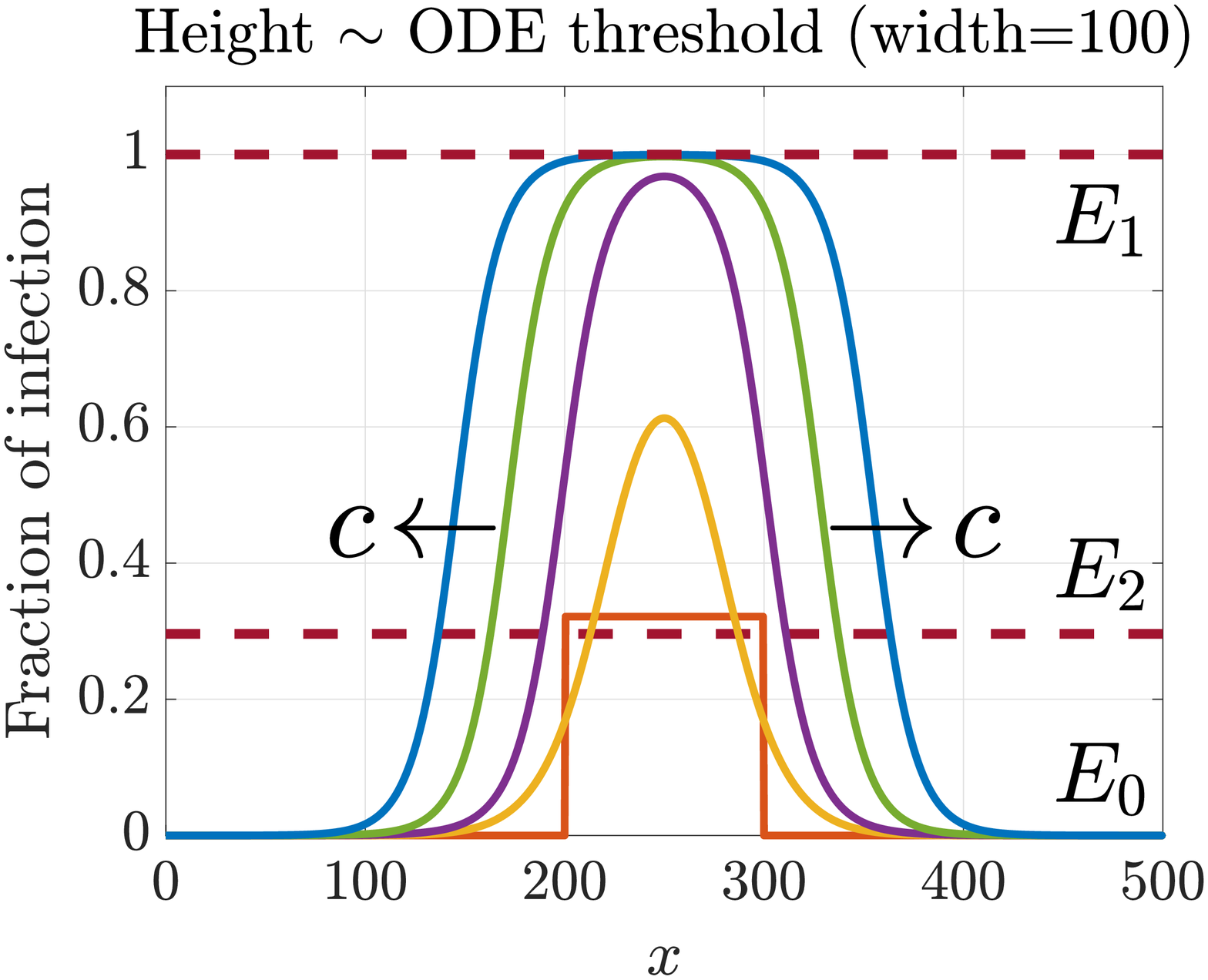}\hfill \includegraphics[width=0.49\textwidth]{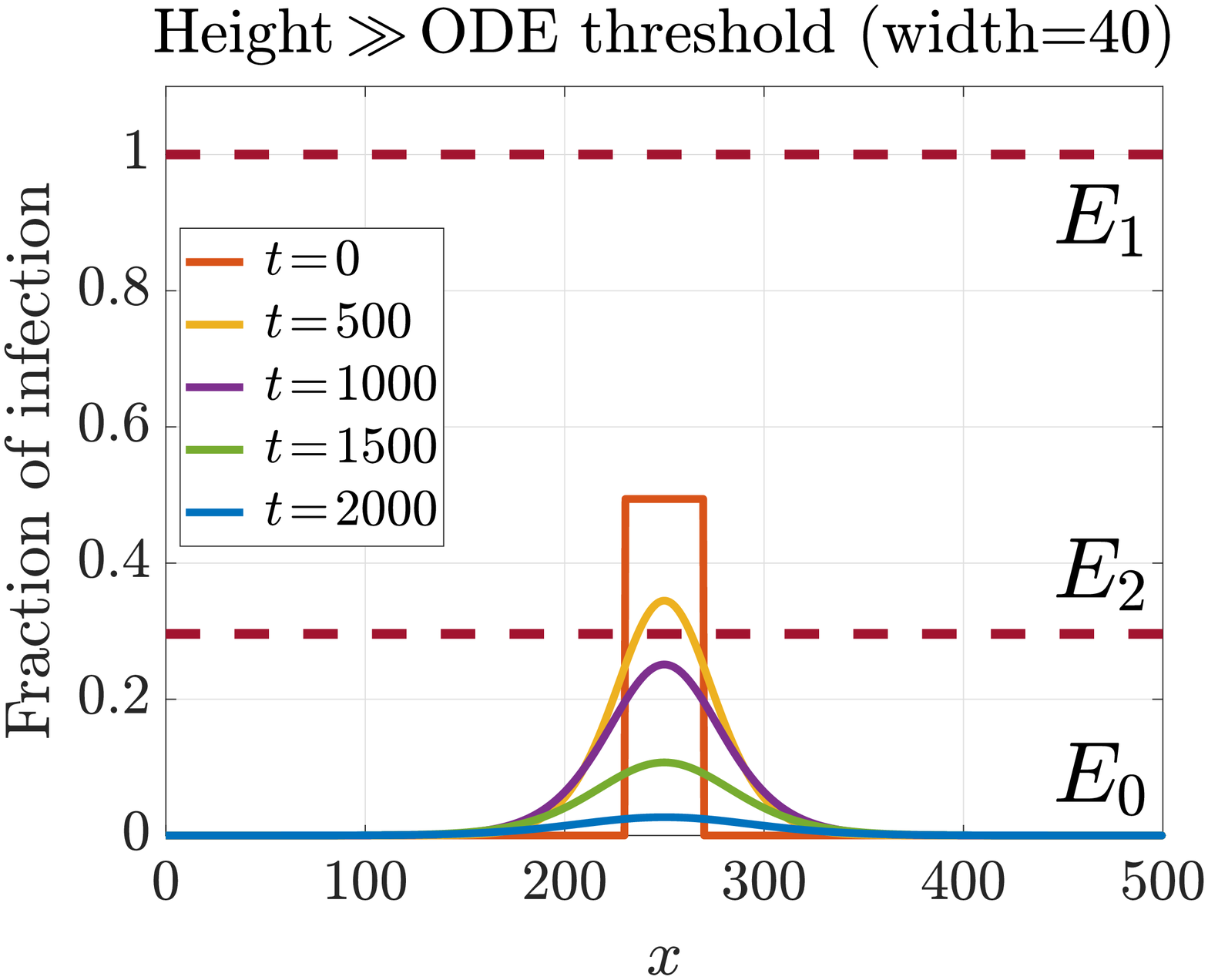}
\caption{Threshold condition for \W invasion depends on the spatial dynamics. Left: invasion happens, and the initial infection step is slightly above the ODE threshold $E_2$. The invasion process takes two stages: wave initiation and propagation. Right: for a much higher initial infection with a narrower step width, the infection collapses. Note: left and right plots share the same legend. \label{fig:PDEODE_twostages}}
\end{figure}

When the infected mosquitoes are released into a field of \Wns-free mosquitoes, the invasion dynamics depend on the competition between the two mosquito cohorts. Typically, given a compact initial condition, a successful invasion happens in two stages (\cref{fig:PDEODE_twostages} left): the wave initiation and wave propagation. 

During the wave initiation stage (\cref{fig:PDEODE_twostages} left, $0<t<1000$), the released infected mosquitoes compete with the native uninfected mosquitoes near the release center. If the initial condition is above the threshold infection level, then the infection wave gradually grows until it reaches the stable high-infection steady states, $E_1$. It is critical to quantify this threshold condition to inform the design of the field trials.

Threshold conditions estimated have been established for the ODE models under the idealized setting where the infected and uninfected mosquitoes are homogeneously mixed. However, in the practical field releases, there is heterogeneous mixing between the two cohorts due to the influx of the released infected population. Simple numerical simulations, as shown in \cref{fig:PDEODE_twostages},  demonstrate that the threshold condition depends on the spatial dynamics, and the threshold condition identified by the ODE models fails to handle the practical field release scenarios. We will focus on identifying the threshold condition with spatial heterogeneity and explore the optimal strategy to establish such an invasion wave.

Once the infection wave has been established, it converges to a traveling wave (\cref{fig:PDEODE_twostages} left, $t>1000$), which joins the stable steady states $E_0$ and $E_1$ and propagates outward with speed $c$. We will characterize the traveling wave solution of the proposed spatial model.

\section{Approximating the 2-PDE model with 1-PDE model\label{sec:1-PDE}}
The complex nonlinear birth term (a rational polynomial factor) in the 2-PDE system \cref{eq:2PDE} makes it difficult to analyze the threshold conditions and traveling wave. We further reduce the 2-PDE model to a 1-PDE approximation for a more manageable analysis. The knowledge gained from the analytical study of the 1-PDE model will provide a reference for the numerical study of the 2-PDE system. Our later investigation indicates that the two models closely resemble each other in various aspects of interest. 

To reduce the number of variables, we introduce $p=v/(u+v)$, the fraction of infection, and look for a differential equation that has the following bistable structure on the right-hand side,
\begin{equation}
p_t \sim p(p-p_2)(1-p)\,,\quad \text{where}\quad  p_2 = \frac{v_2}{u_2+v_2} =\frac{d-a}{d-a+ad}
\label{eq:cubic}
\end{equation}
corresponds to the unstable steady state $E_2$. This is a similar formulation as the cubic approximation in \cite[equation (3)]{barton2011spatial}. To this end, we consider the following transformation
\begin{equation}
u+v = 1-\frac{bd}{a}+\varepsilon\,, \quad  \frac{v}{u+v} = p\,,\label{eq:trans2}
\end{equation}
and write $u=u(p,\varepsilon)$ and $v=v(p,\varepsilon)$. Note that $\varepsilon =0$ at the unstable steady state $E_2$ and the stable high-endemic steady state $E_1$. 
When simulating the field release, where the total mosquitoes population, $u+v$, is near its maximum sustainable size, $\varepsilon$ is a small quantity with little spatial variation. Hence, we follow the idea of asymptotic expansion and approximate the system \cref{eq:2PDE} in terms of this small quantity.

Expanding $p_t = (v/(u+v))_t$ and replacing the time derivatives using the model \cref{eq:2PDE}, upon the parameter transformation \cref{eq:trans2}, we have $p_t = \mathcal{F}(p,\varepsilon,p_{xx},\varepsilon_{xx})$. We then expand the right-hand side at $\varepsilon=0$ and assume $\varepsilon_{xx}\approx 0$, and the O(1) term in the expansion gives
\begin{equation}
p_t = \frac{b(d-a+ad)}{a+a(d-1)p}\,p\,(p-p_2)(1-p)+\Big(D+(1-D)\,p\Big)p_{xx}\,,
\label{eq:1-PDE-general}
\end{equation}
which has a density-dependent diffusion coefficient. The first rational polynomial remains positive around the baseline, and it is an extra factor, comparing to the cubic formulation \cref{eq:cubic}. 

When the diffusion ratio $D=D_2/D_1=1$, that is the same diffusion coefficient for the infected and uninfected females, the equation is reduced to 
\begin{equation}
p_t = \frac{b(d-a+ad)}{a+a(d-1)p}\,p\,(p-p_2)(1-p)+p_{xx}\,.
\label{eq:1-PDE}
\end{equation}

\section{Threshold condition for \W invasion\label{sec:threshold}} 
The threshold condition determines when introducing \Wns-infected mosquitoes will create a sustained infection in the field.

According to the classical results in Fife \cite{fife1979mathematical}, our spatial models are the ``saddle-saddle''-type systems, where the two stable steady states, $E_0$ and $E_1$, are both saddle points in a four-dimensional phase space (see \cref{sec:travel}). It is shown that, for a wide range of initial data $\Phi(x)$, if it satisfies
\begin{equation}\label{eq:threshold1}
\limsup_{x\rightarrow-\infty}\Phi(x)>\alpha, ~~ \liminf_{x\rightarrow\infty}\Phi(x)<\alpha, ~~~\alpha:\text{intermediate unstable equilibrium},
\end{equation}
then the solution uniformly converges to a stable traveling wave \cite[Theorem 4.16 and Corollary 4.18]{fife1979mathematical}. In another word, the ODE threshold state $E_2$ is also a PDE threshold when it's extended to the spatially homogeneous setting. However, condition \cref{eq:threshold1} is not practical for instructing the field releases, as it requires a positive infection present on an infinite domain (as $x\rightarrow -\infty$). We will search for a threshold condition on the initial data which has a compact support.

\subsection{Balanced profiles and critical bubble\label{sec:balance}}
\begin{figure}[htbp]
\includegraphics[width=0.32\textwidth]{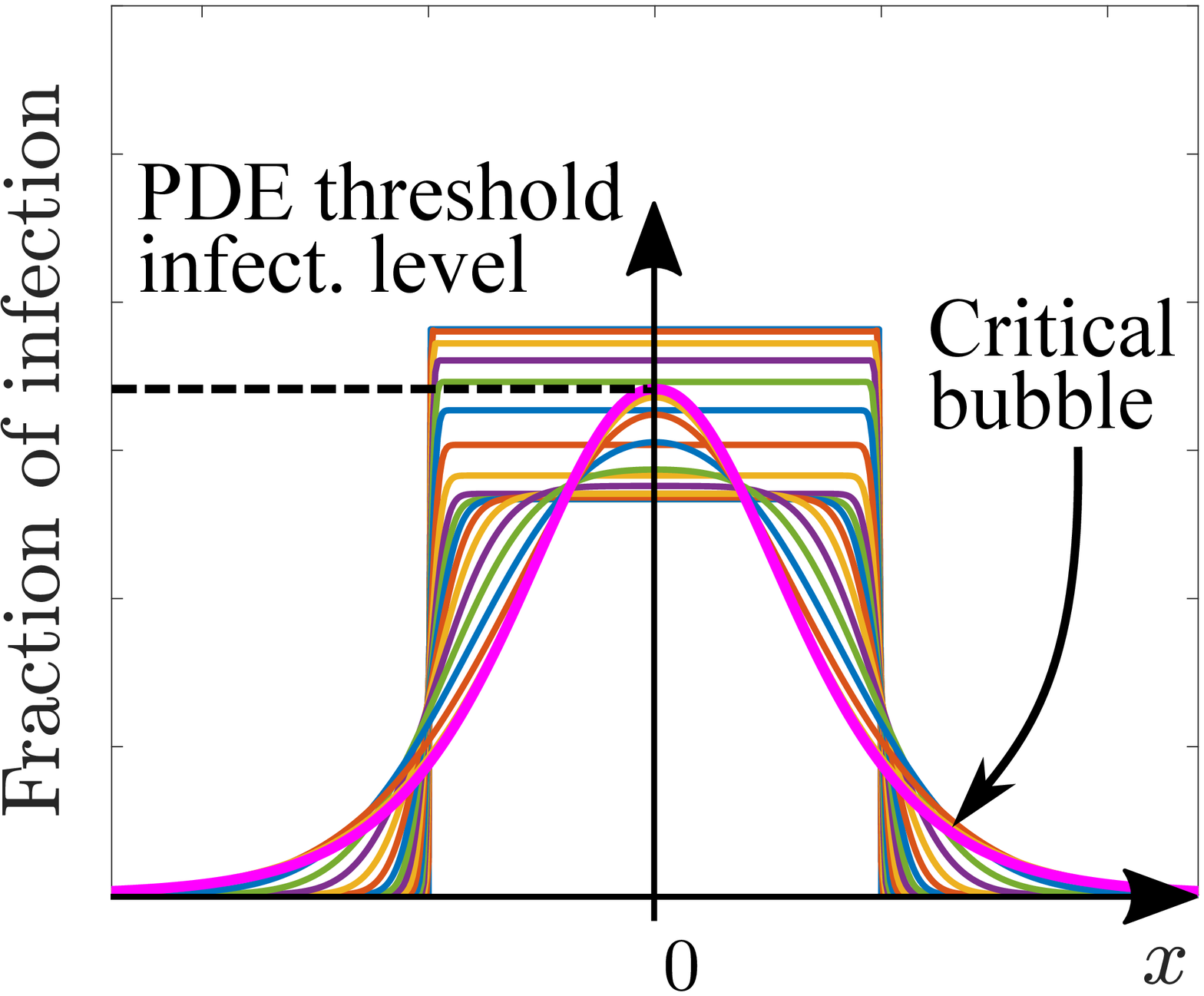}\hfill\includegraphics[width=0.32\textwidth]{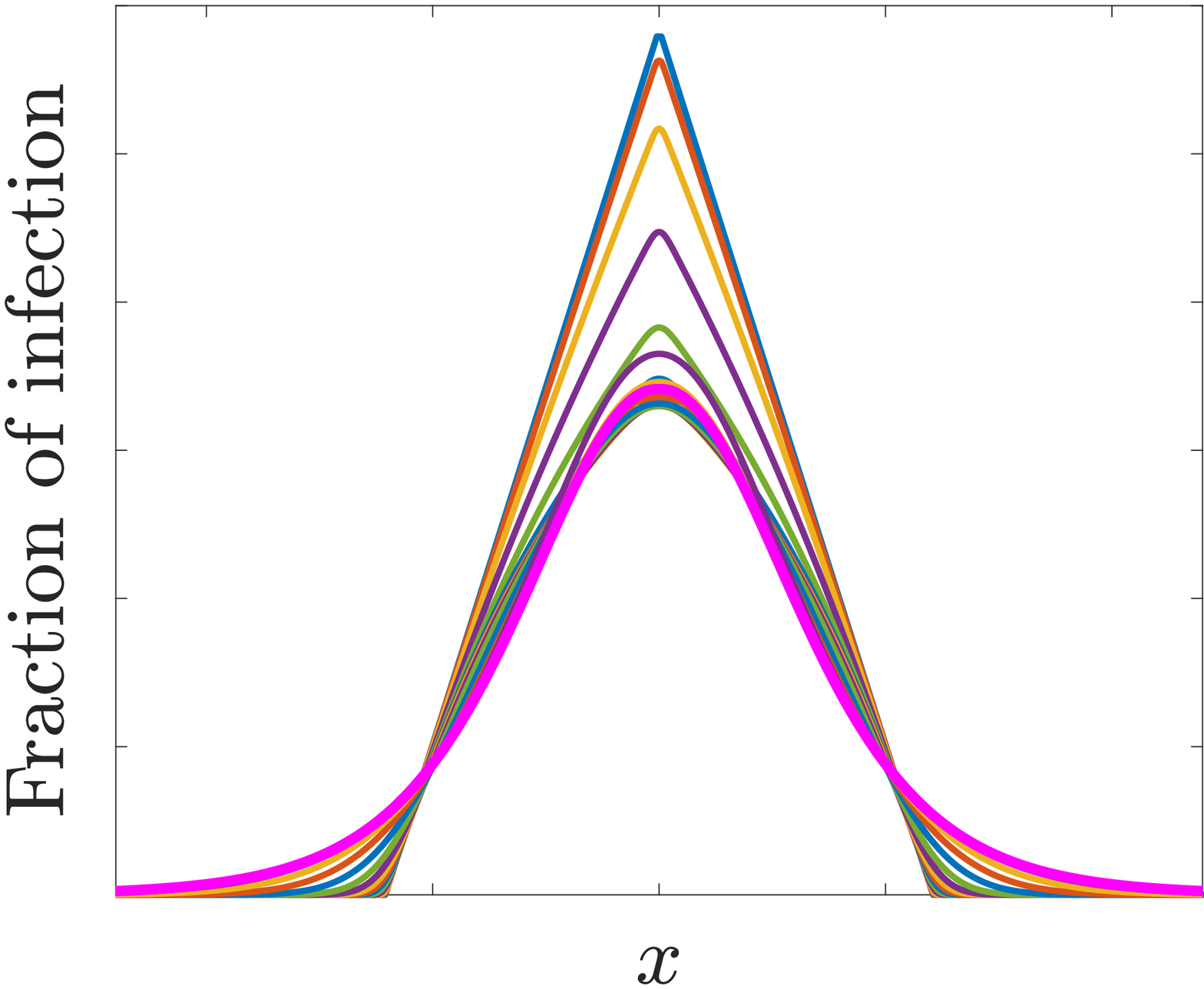}\hfill\includegraphics[width=0.32\textwidth]{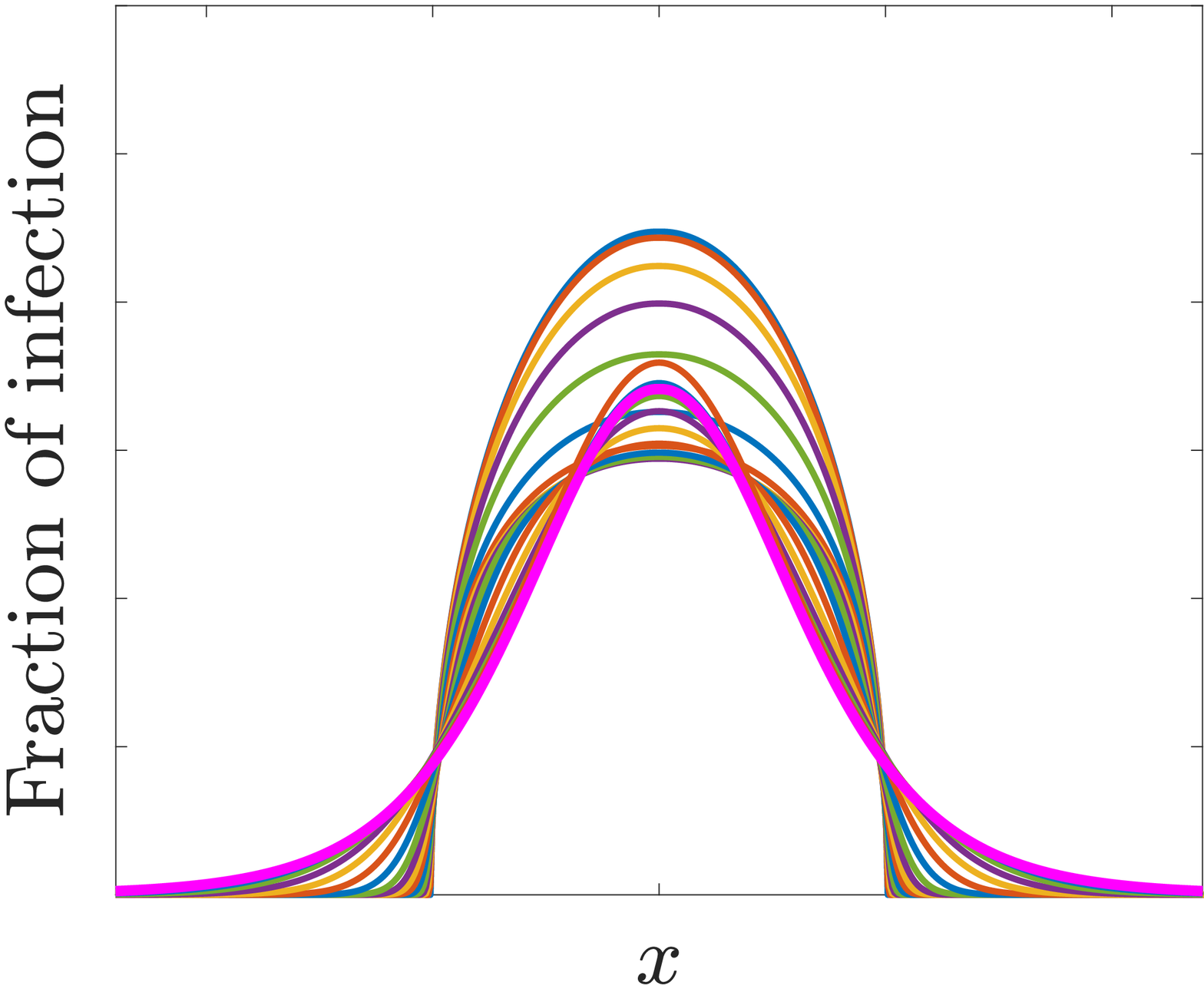}
\caption{Evolution of different initial infection distributions, identified at the corresponding threshold levels, to the same magenta balanced bubble-shaped profile. The final curve, called ``critical bubble'', is the threshold for the PDE model, and the peak is denoted as the PDE threshold level.\label{fig:balanced}}
\end{figure}

For different spatial profiles of release, such as step, triangle, or ellipse (see \cref{fig:balanced}), we can identify the corresponding threshold condition, parameterized by its infection level at the peak. After a short transition period, the threshold profiles all evolve to the same bubble-shaped profile. This unique shape balances the competition of the forces between the growth of infection from reproduction (reaction term) and the spread of the infection from the mosquito diffusion. Rather than attempting to quantify the threshold conditions for an arbitrarily shaped distribution of initially infected mosquitoes, we focus on quantifying the threshold for this balanced bubble-shaped profile.   

We denote the balanced profile at its threshold height (peak at the release center) as the PDE threshold infection level (\cref{fig:balanced} left), and we call the corresponding distribution curve as a \emph{critical bubble}, following the notion in Barton and Turelli \cite{barton2011spatial}. This critical bubble curve is a nontrivial unstable equilibrium. 

By symmetry, in the rest of the paper, we consider only the half-infinite domain (the positive x-axis), and we impose a symmetric boundary condition at $x=0$, which corresponds to the release center.

\subsection{Determining the threshold conditions}
We first analyze the threshold condition for the 1-PDE model \cref{eq:1-PDE-general}. We then numerically study the threshold conditions for the 2-PDE system and compare the results obtained using the two models. 

\subsubsection{Analytical study of the 1-PDE threshold\label{sec:threshold1}}
\paragraph[For D=1]{For $D=1$}
The critical bubble, $p(x)$, is the nontrivial steady state of the boundary value proble \cref{eq:1-PDE}, 
\begin{equation}
p''+h(p)=0\,,
\label{eq:1-PDE-ODE}
\end{equation}
with the boundary conditions
\begin{equation}
p'(0)=p'(\infty)=0\,.
\label{eq:bounary}
\end{equation}
The primes denote the derivative with respect to the $x$, and the nonlinear function $h(p)$ is defined as  
\begin{equation}
h(p)=\frac{b(d-a+ad)}{a+a(d-1)p}\,p\,(p-p_2)(1-p)\,.
\label{eq:hp}
\end{equation}
We multiply both sides of \cref{eq:1-PDE-ODE} by $p'(x)$ and integrate on the $x-$domain $[x_0,\infty]$,
\begin{equation*}
\int_{x_0}^\infty p'(x)p''(x)dx + \int_{x_0}^\infty p'(x)h(p)dx=0\,.
\end{equation*}
Denote $p(x_0)=p_0$, and the last equation can be simplified as
\begin{equation*}
\frac{1}{2}\Big(p'(x)\Big)^2~\Big|_{x_0}^\infty+\int_{p_0}^0 h(p) dp=0\,,
\end{equation*}
which can be rewritten as 
\begin{equation}
p'(x_0)=-\left(-2\int_0^{p_0} h(p) dp\right)^{1/2}
\label{eq:bubble_ODE}
\end{equation}
using the boundary condition \cref{eq:bounary}. Note that $p(x)$ is a decreasing function in $x$.

We set the release center of critical bubble at $x=0$, and $p'(0)=0$. The peak of the critical bubble is the threshold infection level, $p^*=p(0)$. Setting the right-hand side of \cref{eq:bubble_ODE} to be zero, $p^*$ is the root for the nonlinear equation
\begin{equation}\label{eq:Fp} 
\begin{aligned}
&H(p)=\int_0^{p} h(y) dy \\
&=-\frac{b}{6 a (d-1)^4} \bigg((d-1) p \Big(2 (d-1)^2 p^2 (a (d-1)+d) \\
&-3 (d-1) p \left(a (d-1)^2+(2 d-1) d\right)+6 d^3\Big)-6 d^3 \log ((d-1) p+1)\bigg)=0\,.
\end{aligned}
\end{equation}
To derive the shape of the critical bubble, we start from \cref{eq:bubble_ODE} and search for the nontrivial solution for the initial value problem 
\begin{equation}
p'(x)=-\Big(-2H(p)\Big)^{1/2},\quad p(0)=p^*\,,
\label{eq:IVP}
\end{equation}
where $H(p)$ is given in \cref{eq:Fp}. 

\paragraph[For D not equal to 1]{For $D\neq 1$}
The analysis above could be extended for the case when $D\neq 1$, that is we want to find the a nontrivial steady state for \cref{eq:1-PDE-general}:
\begin{equation*}
(D+(1-D)p)p''+h(p)=0\,,
\end{equation*}
with the same boundary condition \cref{eq:bounary}, and $h(p)$ is defined as in \cref{eq:hp}. After normalizing the leading coefficient, we have
\begin{equation*}
p''+h_D(p)=0,\quad h_D(p)=h(p)/(D+(1-D)p)\,,
\end{equation*}
and the rest of the analysis is identical to the $D=1$ case except substituting $h(p)$ with $h_D(p)$. The threshold value, $p^*_{D}$, is the root for the nonlinear equation
\begin{align}
H_D(p)=\int_0^{p} &h_D(y) dy=\nonumber\\
\bigg((d-1) \Big(&(1-D) p (1-d D) \big(a (d-1)^2 (2-(1-D) p)+d (d (-D (2-p)-p+4)\nonumber\\
+&(1-D) p-2)\big)+2 (d-1)^2 D (d-a (1-d D)) \log \left(1+(1/D-1)p\right)\Big)\label{eq:HD}\\
-&2 d^3 (1-D)^3 \log ((d-1) p+1)\bigg) \frac{b}{2 a (d-1)^3 (1-D)^3 (1-d D)}=0\,.\nonumber
\end{align}
The critical bubble satisfies the initial value problem
\begin{equation}
p'(x)=-\Big(-2H_D(p)\Big)^{1/2},\quad p(0)=p_D^*\,.
\label{eq:IVP2}
\end{equation}

The analytical solution for the root of the nonlinear equations \cref{eq:Fp,eq:HD} and the initial value problems \cref{eq:IVP,eq:IVP2} are not available, but they can be numerically solved using simple numerical methods. The \cref{fig:threshold_comp} in \cref{sec:compare} shows the critical bubbles for a range of $D$ values.

\subsubsection{Numerical study of the 2-PDE threshold\label{sec:threshold2}}
To capture the critical bubble for the 2-PDE model \cref{eq:2PDE}, we simulate a continuous point-release strategy, which generates the balanced bubble-shaped profile as discussed in \cref{sec:balance}. We then iterate on different infection levels at the release center, the height of the bubble, to find its threshold level.  We describe the iteration algorithm as follows.

\textbf{Step 1: Point-release to establish balanced profile.} We construct the balanced profile by simulating a point-release process. At time $t = 0$, we release infected mosquito at a point ($x=0$) to the disease-free steady state, that is
\begin{equation}
v(x,0) = \left\{\!
\begin{array}{ll} 
\tilde{v}, & x=0\\
 0, & x\in (0,L]
\end{array}\right.\!\!, \quad u(x,0) = u_0\,, ~~ x\in [0,L]\,.
\label{eq:IC}
\end{equation}
This gives an infection level of $\tilde{p}=\tilde{v}/(u_0+\tilde{v})$ at the release center, and it's referred to as the target infection level. The computational domain $[0,L]$ is sufficiently large such that it allows a natural decay of infection to zero near the right boundary. At $x=0$, we impose the symmetric boundary conditions for $u$ and $v$, and at $x=L$, we allow free boundary conditions with zero-order extrapolations. 

\begin{figure}[htbp]
\centering
\includegraphics[width=0.33\textwidth]{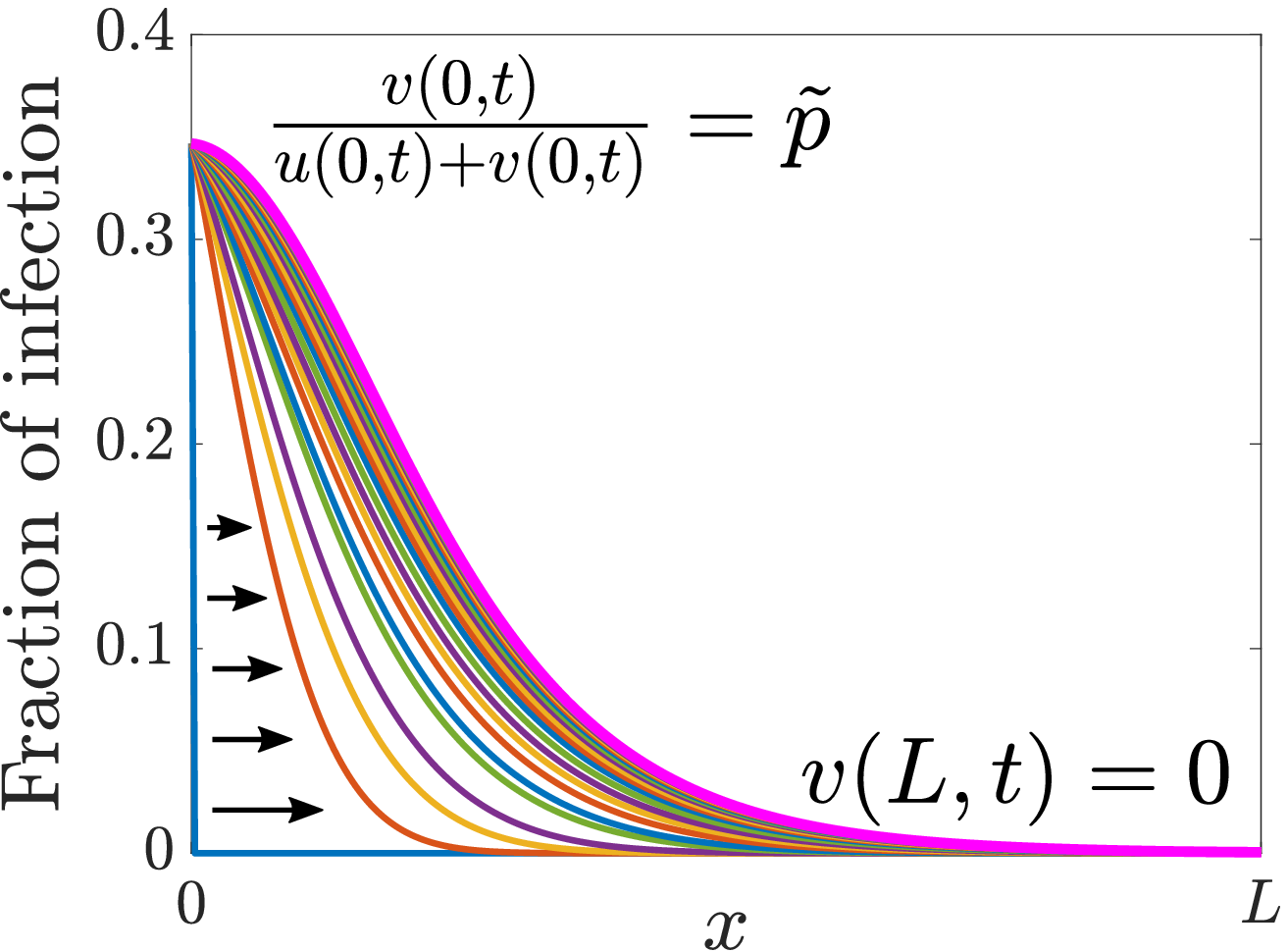}\hfill\includegraphics[width=0.33\textwidth]{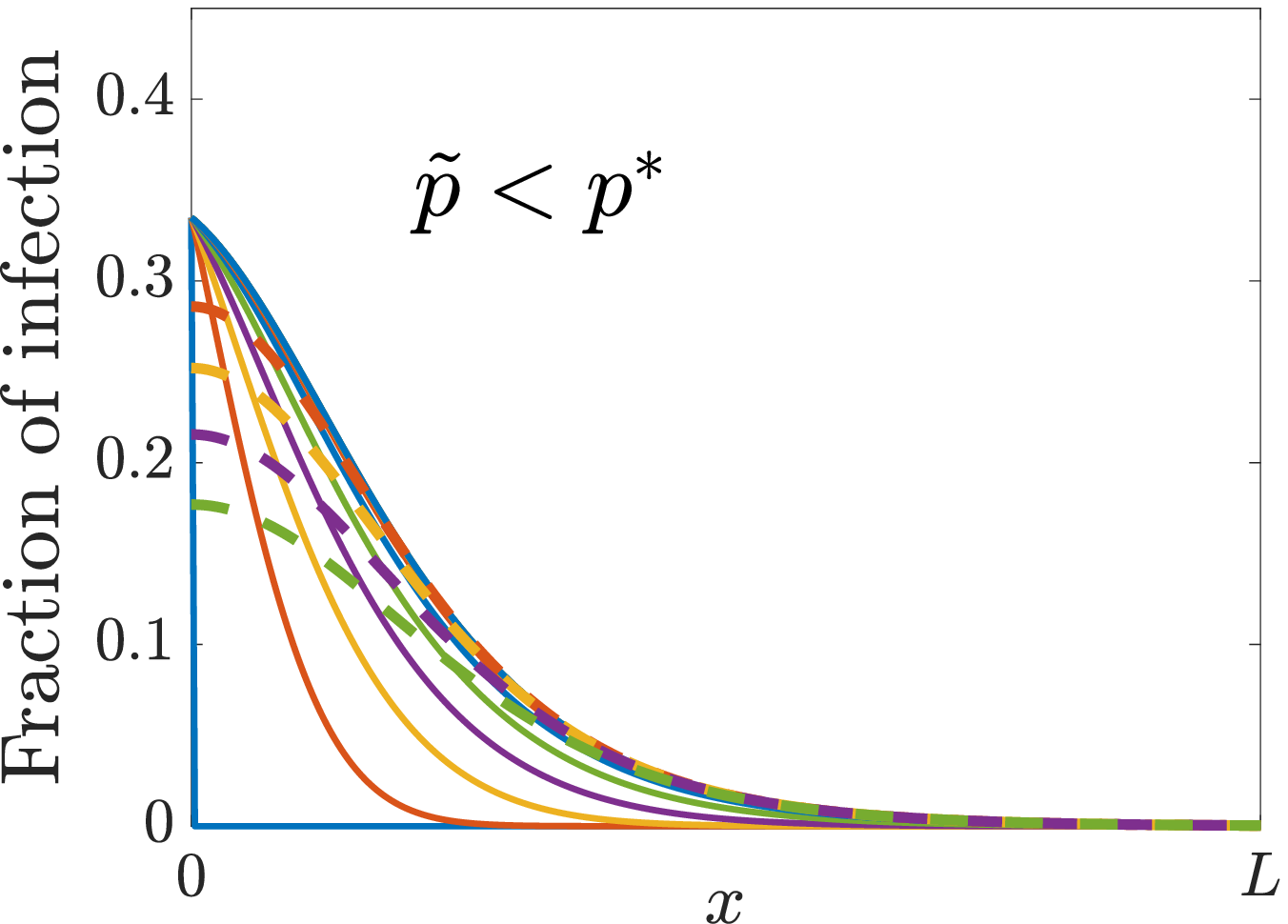}\hfill\includegraphics[width=0.33\textwidth]{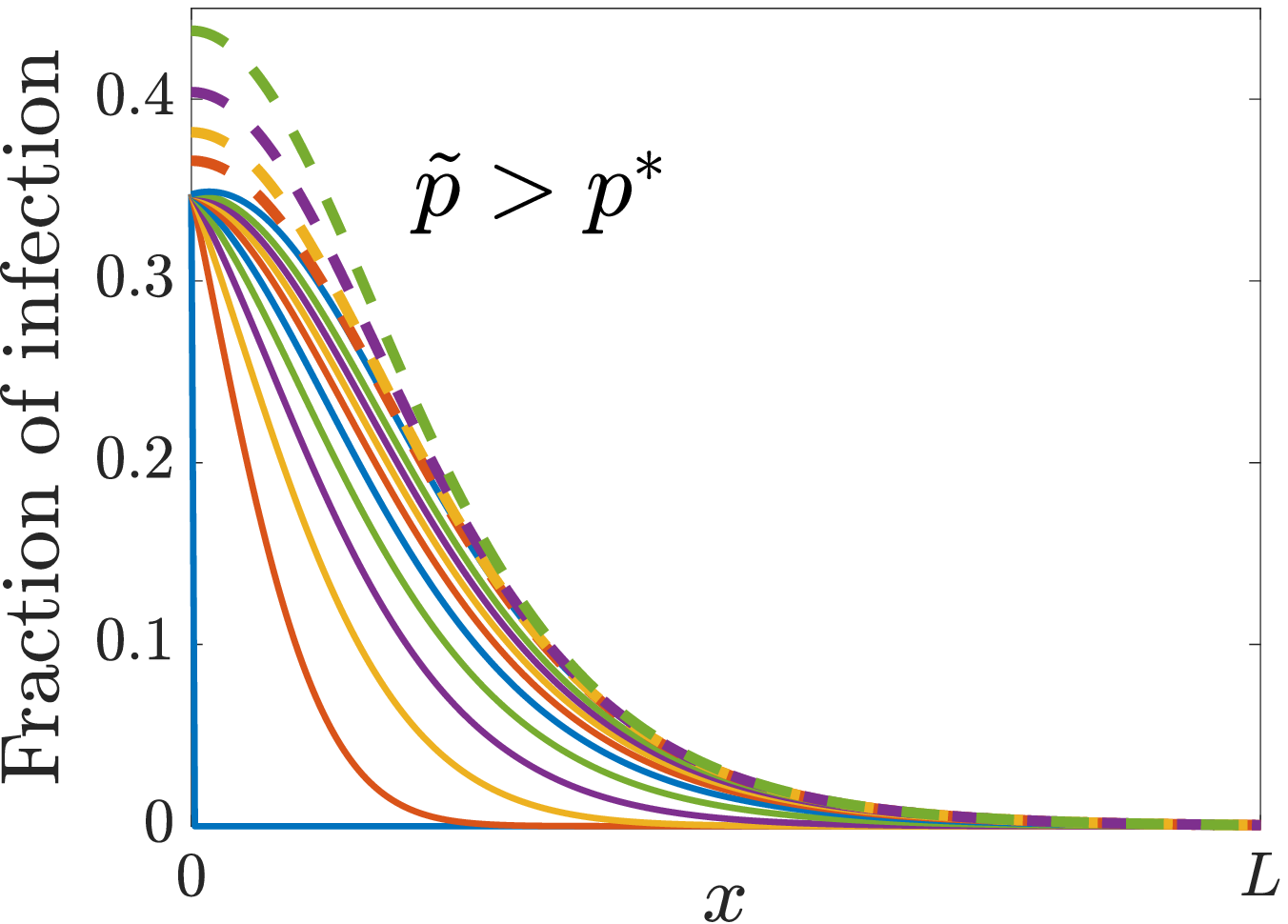}
\caption{Illustration of 2-PDE algorithm. Left: a balanced profile is formulated through a point-release process, where infected mosquitoes are introduced as needed to maintain the target infection level $\tilde{p}$ at the release center $x=0$. Middle \& right: Infection collapses or grows when $\tilde{p}$ is below or above the threshold $p^*$ when stop releasing (dashed curves). \label{fig:step1}}
\end{figure}

When $t>0$, we maintain the target infection level $\tilde{p}$ at the release center by continuously releasing infected mosquitoes there as needed, that is we make the boundary corrections on $v$,
\begin{equation}
v(0,t)=\frac{\tilde{p}}{1-\tilde{p}}\,u(0,t)\,,\quad v(L,t) = 0\,, \quad t> 0\,. \label{eq:fixBC}
\end{equation}
In \cref{fig:step1}, it shows the infection curves of the initial-boundary value problem \cref{eq:2PDE,eq:IC,eq:fixBC} in time, where a balanced profile is established (at time $T_1$) as it reaches a balanced state between the local growth and spatial diffusion.

\textbf{Step 2: Stop releasing.} After the balanced profile is established, we stop the point-release process by removing the boundary correction \cref{eq:fixBC}. We then continue evolving the system with the symmetric boundary conditions on the variables and check the wave front at time $T_2 > T_1$. If the infection collapses, $p(0,T_2)<\tilde{p}$, then it indicates that the target infection level is below the threshold condition ($\tilde{p}<p^*$, \cref{fig:step1} middle); if the infection grows, $p(0,T_2)>\tilde{p}$, then it's above the threshold level ($\tilde{p}>p^*$, \cref{fig:step1} right). 

\textbf{Step 3: Iterate on the target infection level $\tilde{p}$.} We vary the target infection level $\tilde{p}$ and repeat the first two steps until we converge to the threshold level $p^\ast$, where the wave front could maintain its shape after terminating the release. We use a root-finding algorithm, described in \cref{sec:appendix_step3}, to identify this threshold value.



\subsubsection{Comparison of the threshold conditions}\label{sec:compare}
\begin{figure}[ht!]
\centering
\includegraphics[width=0.48\textwidth]{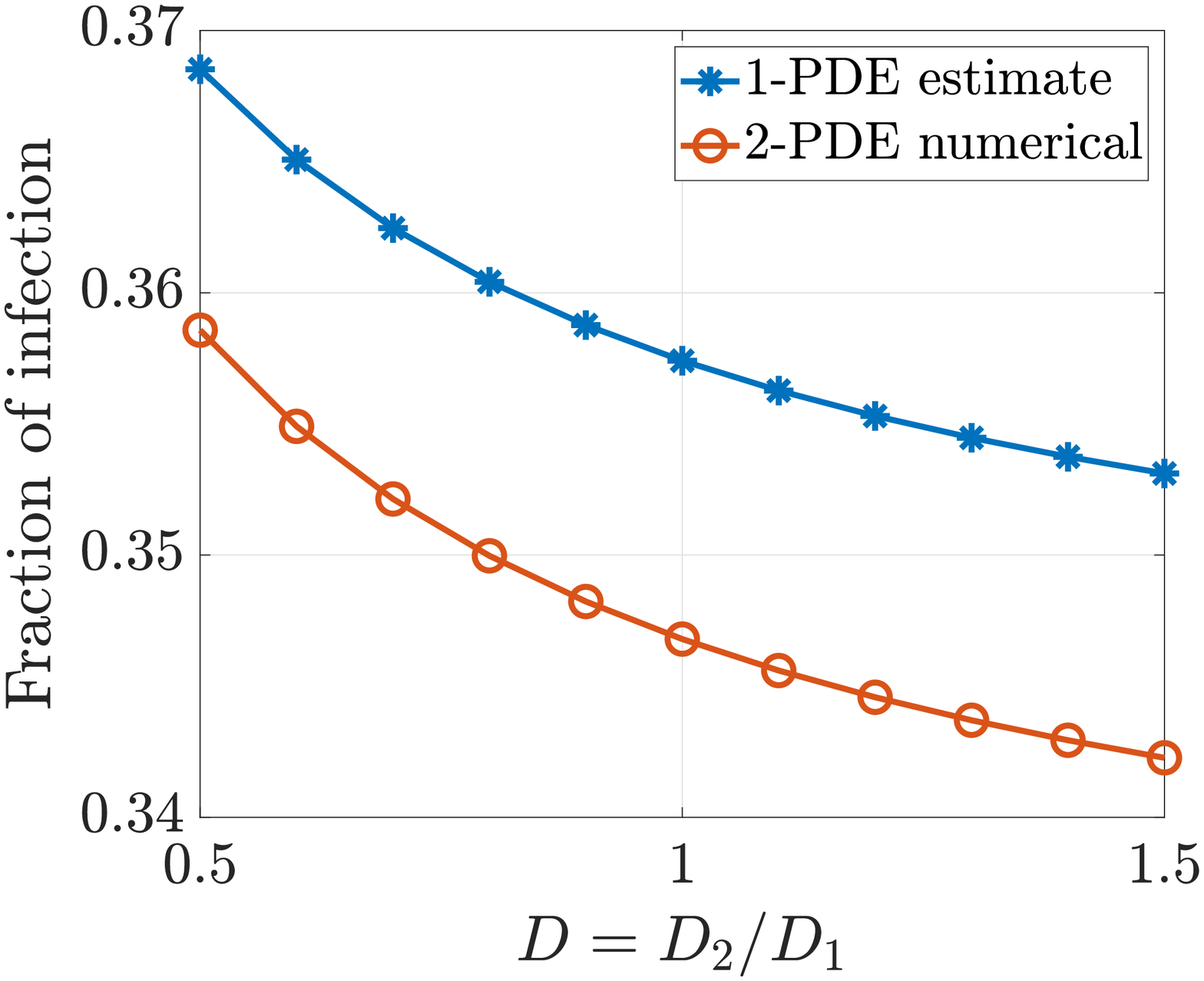}\hfill\includegraphics[width=0.48\textwidth]{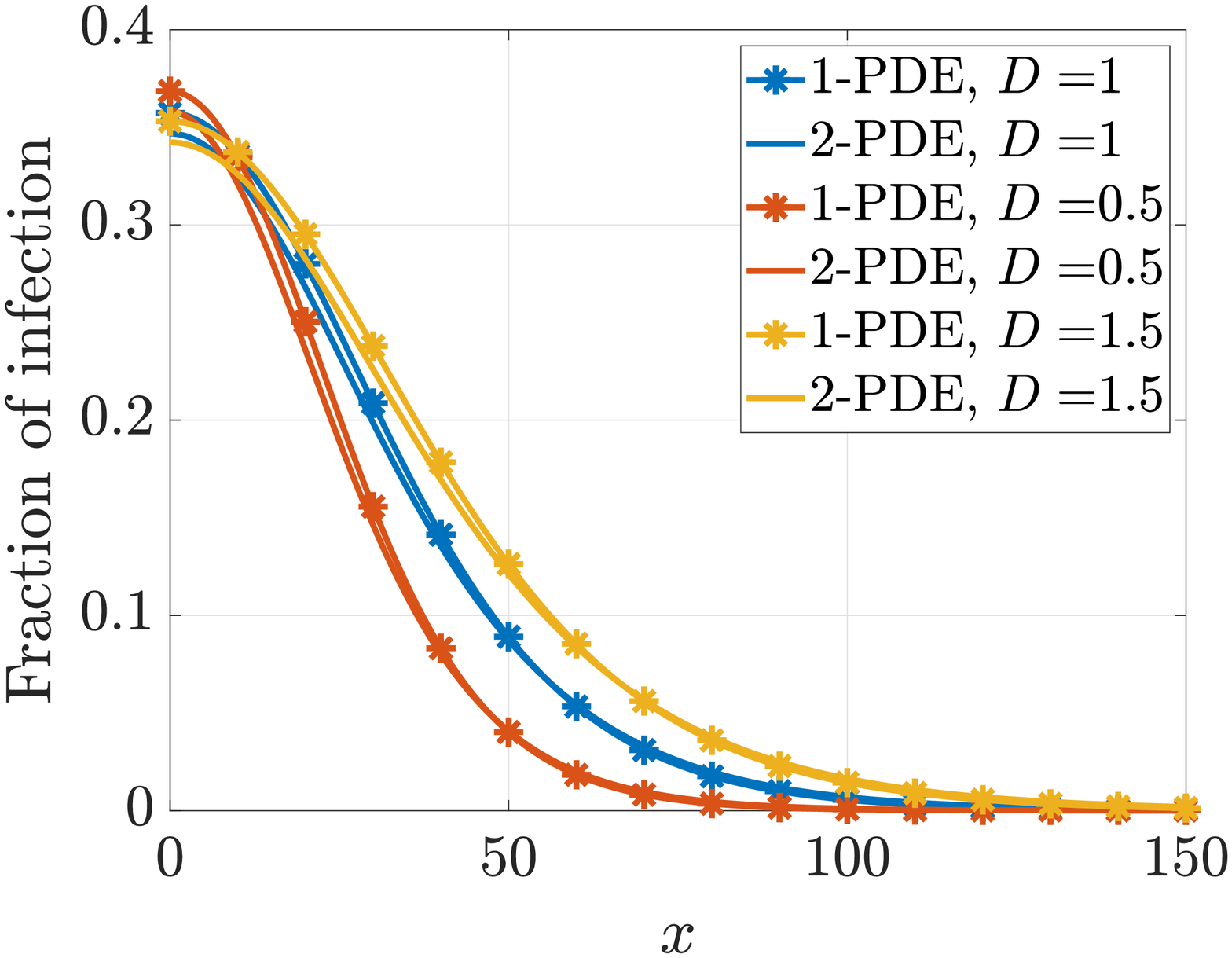}
\caption{Left: Comparison of the estimates for threshold infection levels using the 1-PDE and 2-PDE models. Right: Comparison of the critical bubble shapes using the two models. Overall, the 1-PDE is a good approximation of the 2-PDE model for the threshold conditions. \label{fig:threshold_comp}}
\end{figure}

We compare results of the threshold analysis for the 1-PDE model (described in \cref{sec:threshold1}) and the 2-PDE model (described in \cref{sec:threshold2}).
\paragraph{Threshold infection level} 
We vary the diffusion ratio $D$ and note that the threshold levels for the 1-PDE are slightly larger than the ones for the 2-PDE case (\cref{fig:threshold_comp} left): at the baseline  ($D=1$), the PDE threshold estimates are
\begin{equation*}
p^*_{\text{1-PDE}}\approx0.35741\,, \quad p^*_{\text{2-PDE}}\approx 0.34680\,,\quad p^*_{\text{1-PDE}}-p^*_{\text{2-PDE}}\approx 0.01\,.
\end{equation*}
Moreover, increasing the diffusion ratio $D$ lowers the threshold level for establishing \W infection. This suggests that when the infected mosquito becomes more dispersive ($D_2$ increases), it helps the infection spread out to the nearby region and establish the infection wave front.

Also, all of the PDE threshold levels are above the ODE threshold, determined by the unstable steady state $E_2$. At the baseline values, we have
\begin{equation*}
p^*_{\text{ODE}}=\frac{v_2}{u_2+v_2} = \frac{d-a}{d-a+ad}\approx 0.2284\,.
\end{equation*}
That is, the ODE threshold values can significantly underestimate the infection levels needed, which emphasizes the necessity for incorporating spatial dynamics to give a more reliable prediction for the \W invasion in the field. 

\paragraph{Critical bubble shape}
\Cref{fig:threshold_comp} right compares the 1- and 2-PDE critical bubbles. There is a small discrepancy near the release center, which corresponds to the difference in the threshold infection levels ($\approx 0.01$, shown on the left).  

As the diffusion ratio $D$ increases, the dispersion for infected mosquitoes increases, and the critical bubble becomes wider with a fatter tail when moving towards the edge of releasing region. This may affect the distance between the release locations when there are multiple releasing sites and superposition of the invasion waves happens.

Overall, we see that the 1-PDE analysis gives a good approximation to the 2-PDE model in terms of the threshold-related quantities. Besides, the iterative algorithm for identifying the 2-PDE threshold is much more computationally expensive than the approach taken in the 1-PDE case. Hence, the reduced 1-PDE model is a useful reference that infers insights for the complex 2-PDE model. 

\subsection{Practical considerations for bubble and non-bubble thresholds\label{sec:323}}
When releasing infected mosquitoes in the field, practical considerations such as the total number of mosquitoes released, duration of the release program, and different spatial profiles may be associated with the implementation and cost of the field trials. We here present how these quantities are impacted by the diffusion ratio $D$ during the bubble formulation. We also compare different non-bubble-shaped release profiles and observe that the critical bubble has an optimal shape with a minimal release number.

\subsubsection{Release number for critical bubble establishment\label{sec:release}} We consider the point-release process for the critical bubble establishment (\cref{fig:step1}), where infected mosquitoes are released at one point to maintain the target infection level $\tilde{p}=p^*$. To calculate the total release number during the process, we estimate the (accumulative) released number $R(t)$, 
\begin{align}
R_t &= \frac{d}{dt}\left(\int_0^\infty v(x,t)dt\right)-\int_0^\infty  a (1-u-v) v-b\,d\,v\,dt\,, \label{eq:Rt}\\
& = \int_0^\infty Dv_{xx}(x,t)dt = - D\, v_x(0,t)\,. \label{eq:Rt1}
\end{align}
In \cref{eq:Rt}, the release rate is estimated by the change in the total infected population, excluding the contribution from the mosquito net growth rate. Assume  $v_x(\infty) = 0$, and the release rate depends on the influx of infection from the left boundary. We solve \cref{eq:Rt1} simultaneously with the main model as a diagnostic equation.

\begin{figure}[htbp]
\centering
\includegraphics[width=0.325\textwidth]{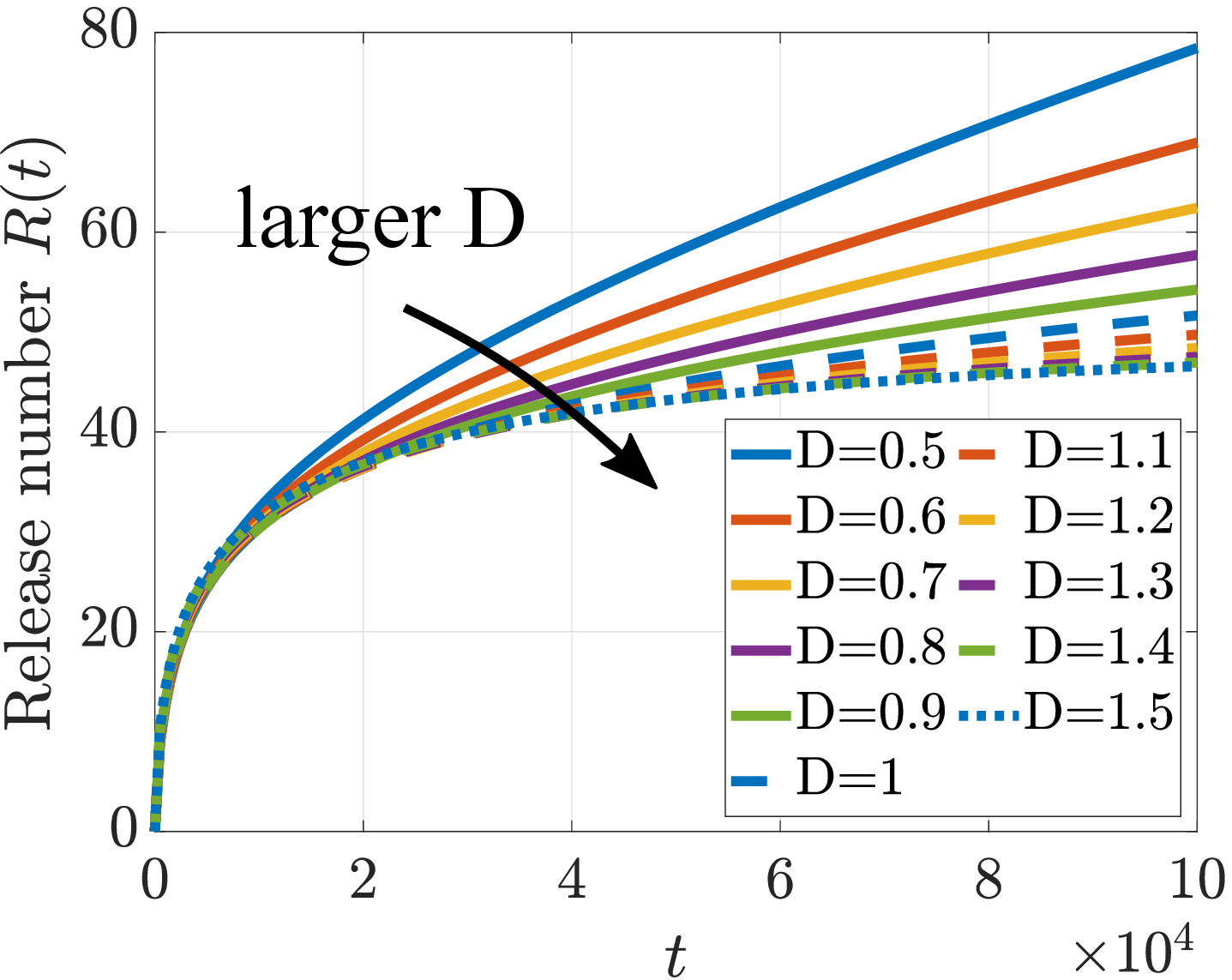}\hfill\includegraphics[width=0.325\textwidth]{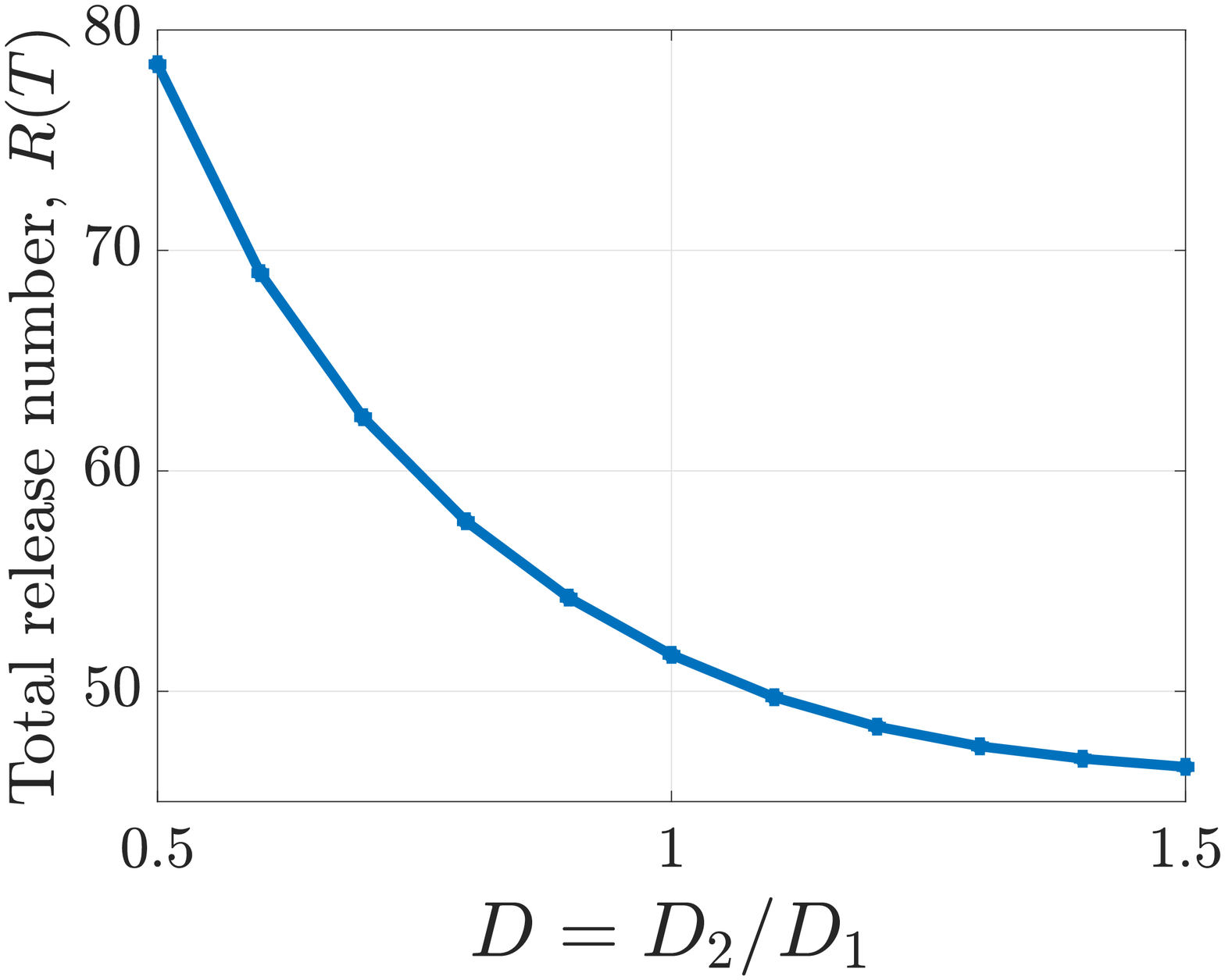}\hfill\includegraphics[width=0.325\textwidth]{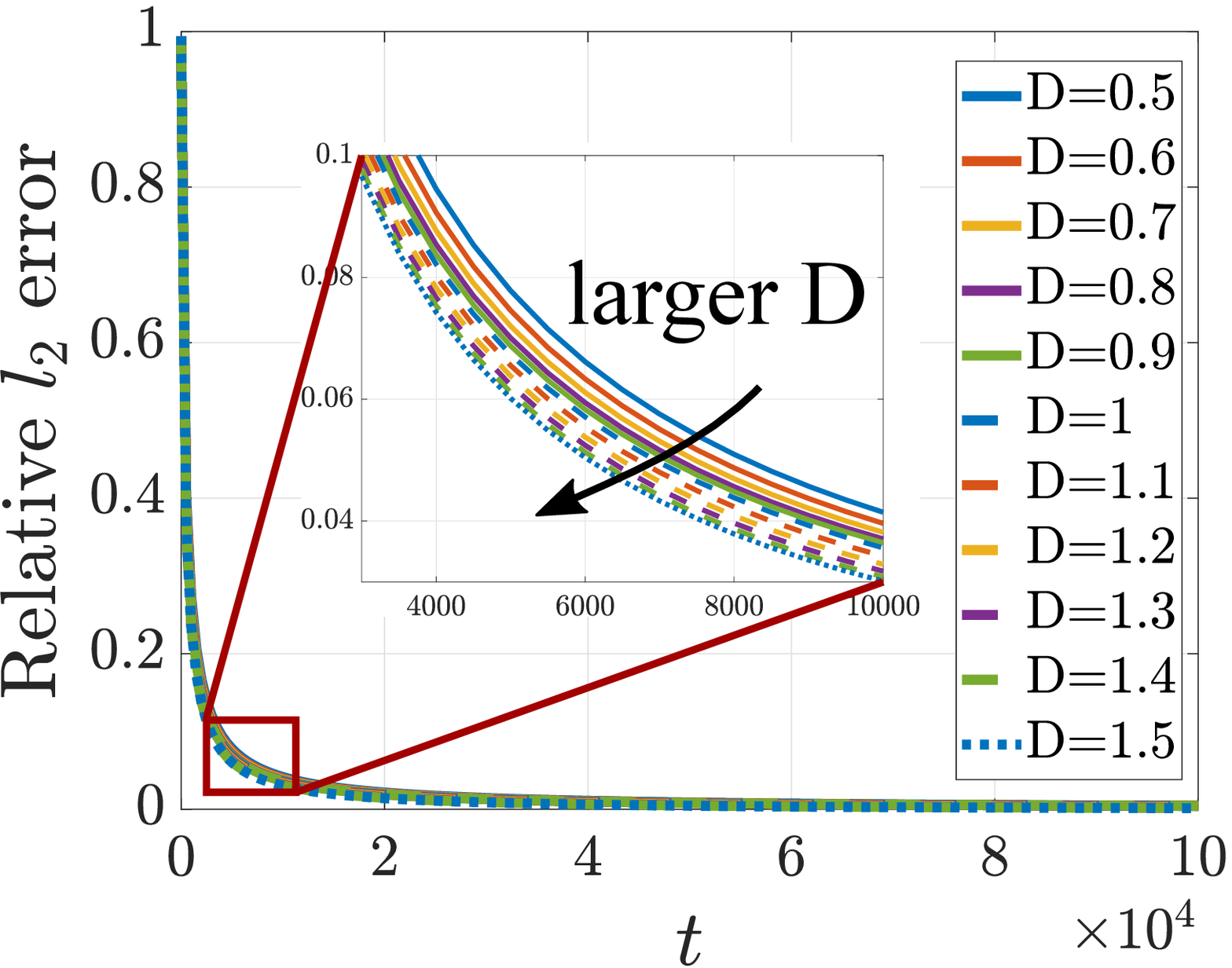}
\caption{Left \& middle: when diffusion ratio $D$ increases, it lowers the total release number needed to establish the critical bubble. Right: For a larger $D$, it have a slightly faster convergence rate, which suggests a faster establishment of the critical bubble. \label{fig:release_D}}
\end{figure}

In \cref{fig:release_D} left, the initial stage of the point-release process requires a large release to maintain the target infection level. As the critical bubble forms, the infection density at the release center becomes more stable, and fewer mosquitoes need to be released each day. Eventually, the release curve reaches a plateau (may take as long as $t=10^6$ for $D=0.5$), where no more infected mosquitoes are released and the established critical bubble can sustain itself in time. As the diffusion ratio $D$ increases (or a faster dispersion of infected mosquitoes $D_2$), fewer infected mosquitoes need to be released before the solution converges (middle plot), and the infection curve converges faster to the critical bubble (right plot). This can also be seen from the release curves (left), where the curves for larger $D$ becomes flat sooner.

\subsubsection{Critical bubble as an optimal spatial threshold profile}
The critical bubble is a balanced spatial configuration of the infection. We can also identify the threshold conditions for unbalanced spatial profiles, such as step, triangle, or ellipse. However, as shown in \cref{fig:balanced}, these threshold profiles evolve to the critical bubble in time. This leads to a natural question: Does the critical bubble represent an optimal infection distribution to give rise to an invasion wave?  To this end, we compare the unbalanced threshold profiles to the critical bubble by measuring the release numbers.

\begin{figure}[htbp]
\centering
\includegraphics[width=0.48\textwidth]{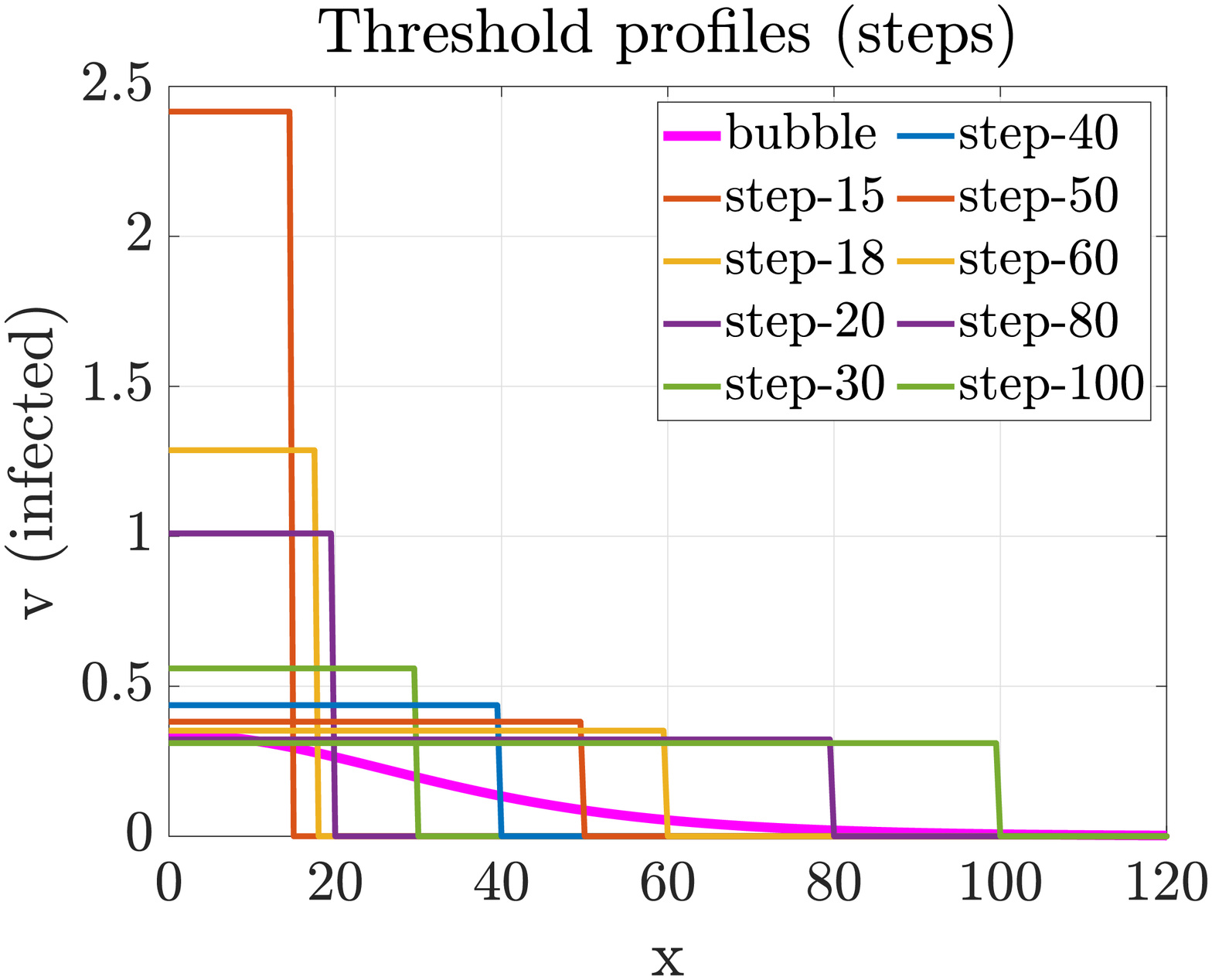}\hfill\includegraphics[width=0.48\textwidth]{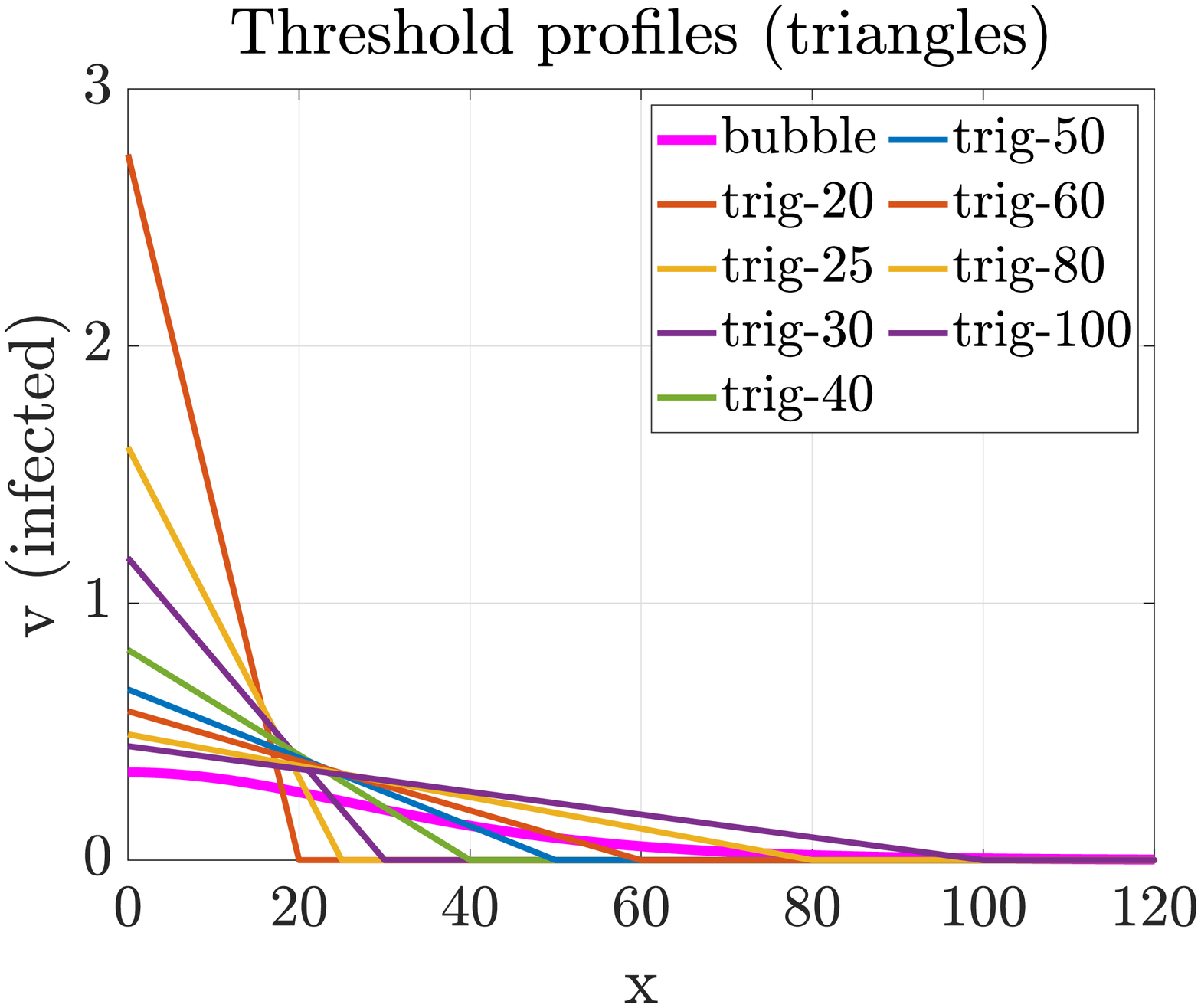}\\ \vspace{3pt}\includegraphics[width=0.48\textwidth]{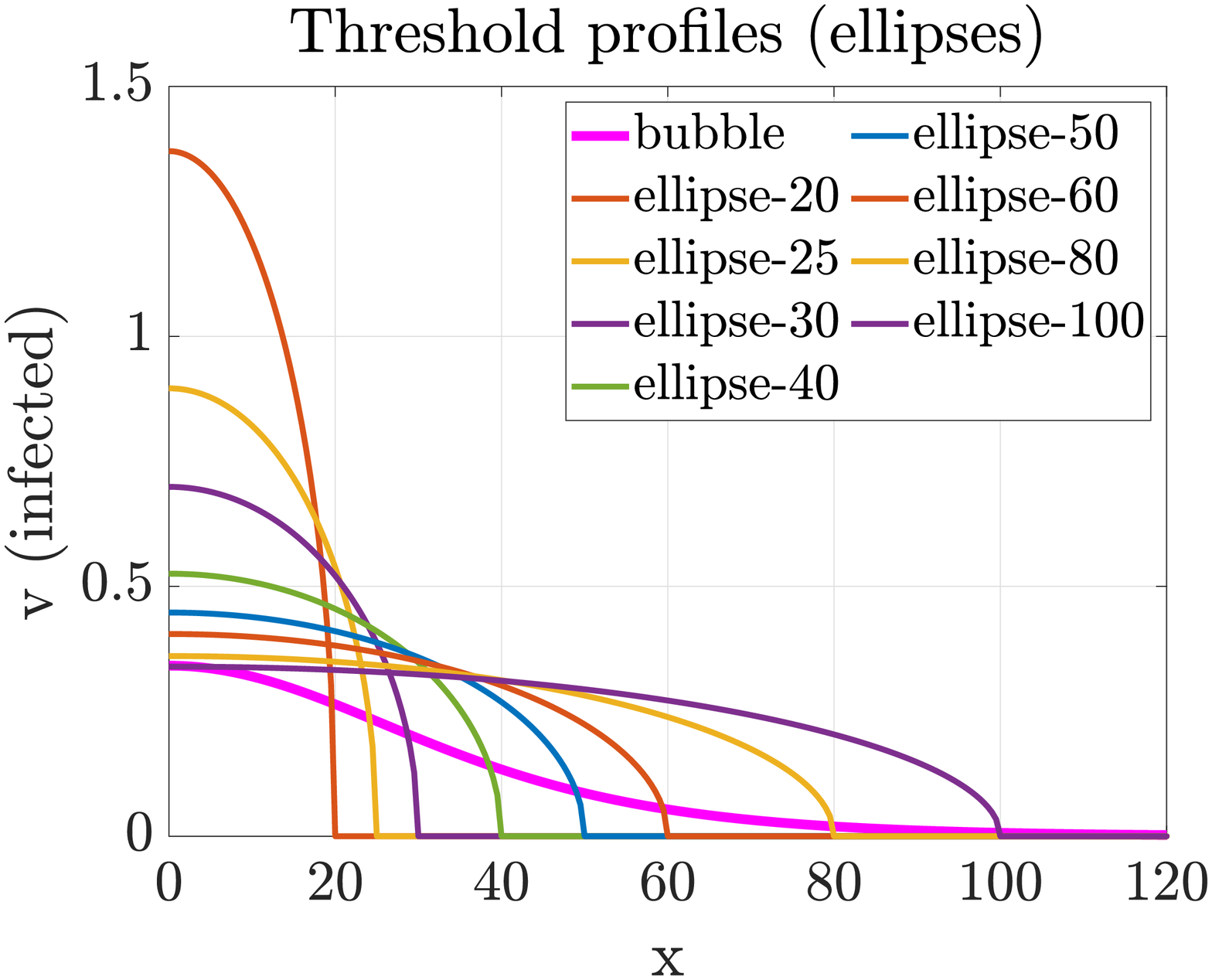}\hfill \includegraphics[width=0.48\textwidth]{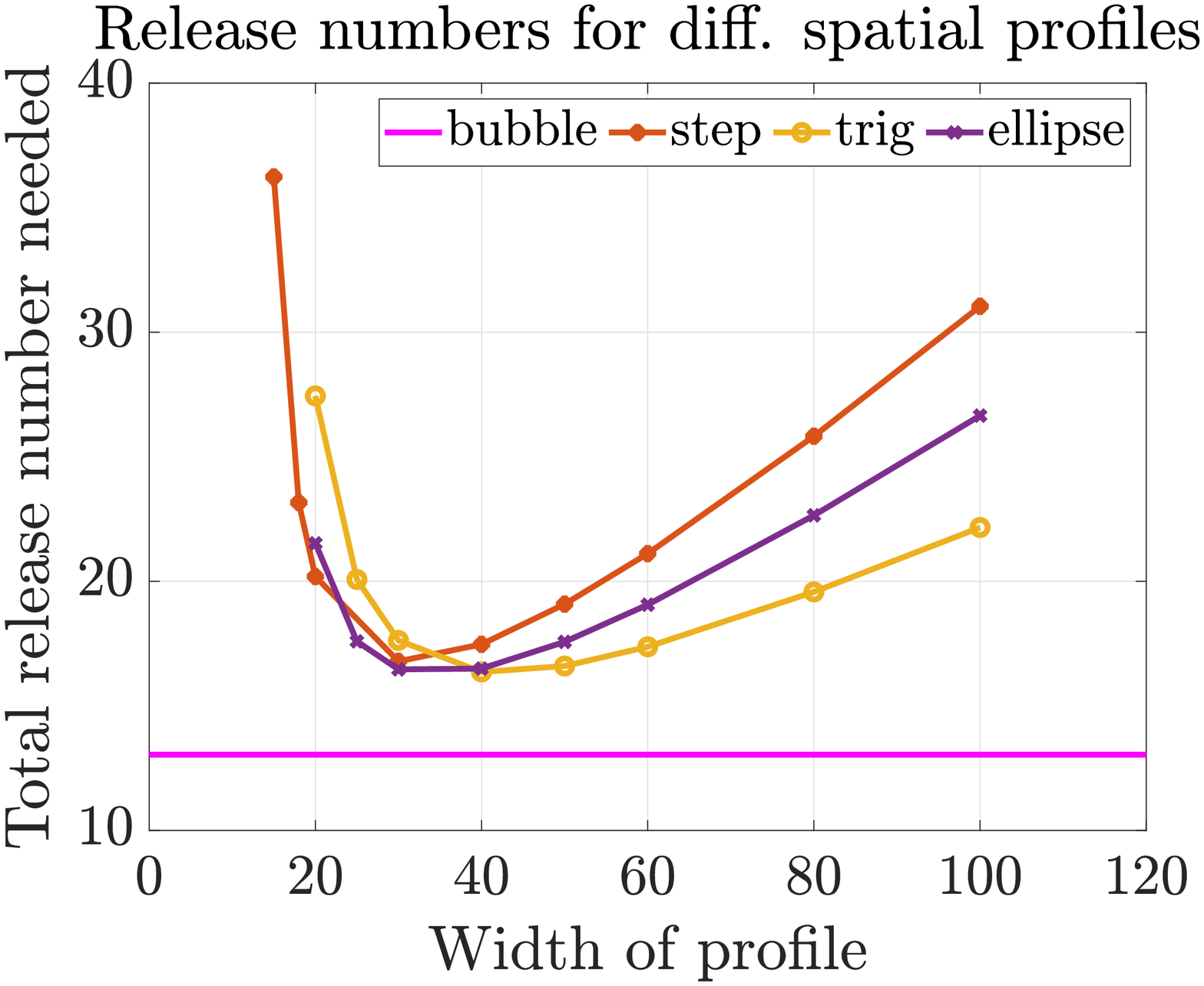}
\caption{Comparison of threshold conditions for unbalanced profiles and critical bubble. \textbf{Top \& bottom left}: threshold curves for step, triangular, elliptical releases with different widths. \textbf{Bottom right}: the total release number, the area under the curve, for different threshold curves. The critical bubble has the smallest infection area, compared to other unbalanced profiles. \label{fig:release_shapes}}
\end{figure}

We first consider the step release profile. For a fixed width of the step, we can find its threshold condition, which is the minimum height needed for invasion (see \cref{fig:release_shapes}, top left). We then calculate the total release number needed as the area under the threshold curve. We note that this corresponds to a different release design from the point-release process described previously, where infected mosquitoes are released continuously at one point to form a bubble-shaped front in time. Here, it assumes that the infected mosquitoes are distributed in a given shape and released all at once. Among all the thresholds curves for different step widths, the optimal step width that has the minimal release number is around $30$ (\cref{fig:release_shapes}, bottom right), and all the step widths require greater release numbers than the critical bubble does.

We then consider two other unbalanced profiles, triangles and ellipses of different widths. Similar to the step case, there is an optimal width (around 40 and 30, respectively) that gives the minimal release number. The release number curves for all the spatial configurations are above the critical-bubble curve.

These results support the observation that the critical bubble has an optimal spatial distribution that requires a fewer amount of infection for establishing wave invasion, compared to the other simple unbalanced distributions we tried. This comes from the advantage of being a balanced profile, where the reaction (net growth) and diffusion (mosquito dispersion) have been balanced at each location. In contrast, for an unbalanced distribution, the infection curve has to go through adjustments before reaching a balanced state due to the competing dynamics. This causes the waste of infection due to the local carrying capacity constraint and natural morality in time. 

We also note that the critical bubble may not be a practical design for the field trials. Unlike the uniform step profile or the point release, the shape of the bubble requires varying the release quantity as a function in space. Nevertheless, the study of the critical bubble serves as a useful theoretical reference. As one can observe from the comparison in \cref{fig:release_shapes}, those spatial configurations that give the minimal release numbers are the ones that more closely mimic the bubble shape in its shape family.  We also caution the reader that these results are for a one-dimensional system, and the shape of the critical bubble will be different for a two-dimensional release pattern.

\section{Traveling wave propagation of \W invasion\label{sec:travel}}
When the released infected mosquitoes are above the threshold, the \W infection can be sustained and the infection wave propagates to the nearby zero-infection region in a traveling wave form. The speed and shape of this traveling wave will be determined by the local environment (model parameters) and are independent of the initial conditions. 
\subsection{Existence of traveling wave solutions} 
We discuss the existence of traveling wave solutions for both the 1-PDE and 2-PDE models.
\subsubsection{Classical results for 1-PDE model}
Consider the reduced 1-PDE model $p_t = h(p)+p_{xx}$, where $h(p)$ is defined in \cref{eq:hp}. The traveling wave solution has the form $p(x,t) = P(x-ct)=P(z)$, and it satisfies the ODE 
\begin{equation}
P''+cP'+h(P)=0\,,\quad \lim_{z\rightarrow-\infty} P(z)=1\,,\quad \lim_{z\rightarrow\infty} P(z)=0\,,
\label{eq:travel}
\end{equation}
where we set the boundary conditions to join the two steady states $P=1$ and $P=0$. We look for a right-going traveling wave ($c>0$) that leads to the invasion and expansion of \W infection. 

Let $\mathbf{X} = [P,W=P']^\top$, the ODE can be rewritten as a system of first-order ODEs
\begin{equation}\label{eq:ODE1}
\mathbf{X}'=\frac{d}{dz}
\begin{bmatrix}
P \\
W \end{bmatrix}
=\begin{bmatrix}
W \\
-cW-h(P) 
\end{bmatrix},
~ P_1=\mathbf{X}(-\infty)=
\begin{bmatrix}
1 \\
0
\end{bmatrix},~
P_0 = \mathbf{X}(\infty)=
\begin{bmatrix}
0 \\
0\end{bmatrix}.
\end{equation}
The traveling wave solution that we are looking for corresponds to a trajectory in the phase plane $(P,W)$ of the system \cref{eq:ODE1}, connecting the two steady states $P_1$ and $P_0$. The existence of such a trajectory depends on the type of the steady states. To this end, we linearize the system and obtain the Jacobian matrix,
\begin{equation*}
\mathcal{J}_p=
\begin{bmatrix}
0 &1 \\
-h'(P) & -c \end{bmatrix}, \quad \text{with eigenvalues}\quad\lambda_{\pm} = \frac{-c\pm\sqrt{c^2-4 h'(P)}}{2}\,.
\end{equation*}
Since $h'(1)=-b<0$ and $h'(0) = b\left(1-d/a\right) <0$
at the baseline, the two eigenvalues are real and have opposite signs around the steady states $P_1$ and $P_0$, which are both saddle points. 

This ``saddle-saddle'' scenario has been discussed thoroughly for a general reaction function $h(p)$: assume there is only one internal zero in $(0,1)$, except for translation in coordinating system, there exists one and only one traveling wave front \cite[Theorem 4.15]{fife1979mathematical}, and the wave front is a stable solution \cite[Corollary 4.18]{fife1979mathematical}. 

\subsubsection{Inference for 2-PDE model}
Following the similar idea as in the 1-PDE model, we shall see that we also have the ``saddle-saddle'' scenario. We present the preliminary steps below and infer that the same conclusions (existence, uniqueness, stability) hold for the 2-PDE model due to the similarity between the two models. However, the rigorous proof for the general two-equation reaction-diffusion system remains an open question to the authors' knowledge.

We look for the traveling wave solution for the 2-PDE model
\begin{equation}
\begin{aligned}\label{eq:ODE4}
&u_t = f(u,v)+u_{xx}\,,\quad f(u,v)=\frac{u}{u+d\,v} (1-u-v)u  -b\,u\,,\\
&v_t = g(u,v)+Dv_{xx}\,,\quad g(u,v)=a (1-u-v) v-b\,d\,v\,,
\end{aligned}
\end{equation}
of the form $u(x,t)=U(x-ct)=U(z)$ and $v(x,t)=V(x-ct)=V(z)$. Substituting the traveling wave form into \cref{eq:ODE4}, we have $U$ and $V$ need to satisfy 
\begin{equation*}
\begin{aligned}
&U''+cU'+f(u,v)=0\,, \quad \lim_{z\rightarrow-\infty} U(z)=0\,,\quad \lim_{z\rightarrow\infty} U(z)=u_0\,,\\
&V''+cV'+g(u,v)=0\,,\quad \lim_{z\rightarrow-\infty} V(z)=v_1\,,\quad \lim_{z\rightarrow\infty} V(z)=0\,,\\
\end{aligned}
\end{equation*}
and we look for a right-going traveling wave ($c>0$). Let $\mathbf{X} = [U,K\!=U',V,Q=V']^\top$, and the system can be rewritten as system of first-order ODEs
\begin{equation}\label{eq:ODE5}
\mathbf{X}'\!=\!
\frac{d}{dz}\!\begin{bmatrix}
U \\
K \\
V \\
Q
\end{bmatrix}
\!=\!\begin{bmatrix}
K \\
-cK-f(U,V) \\
Q \\
-cQ-g(U,V)
\end{bmatrix}\!\!,
~~ E_1\!=\!\mathbf{X}(-\infty)\!=\!
\begin{bmatrix}
0 \\
0\\
v_1 \\
0
\end{bmatrix}\!\!,~~
E_0\!=\!\mathbf{X}(\infty)\!=\!
\begin{bmatrix}
u_0 \\
0\\
0 \\
0
\end{bmatrix}.
\end{equation}

The traveling wave solution corresponds to a trajectory in the phase plane\\ $(U,K,V,Q)$ of the system \cref{eq:ODE5}, connecting the two steady states from $E_1$ to $E_0$ (see \cref{fig:PDEODE_twostages} left). In particular, we look for a physically relevant monotone solution, where $\mathbf{X}$ is increasing in $U$ and decreasing in $V$, and the trajectory should stay within the following domain
\begin{equation*}
0\le U\le 1\,,\quad K>0\,, \quad 0\le V\le 1\,,\quad Q<0\,.
\end{equation*}
To determine the types of stability for the steady states $E_0$ and $E_1$, we linearize the system around them. The linearization of system \cref{eq:ODE5} at $E_0$ gives the Jacobian matrix
\begin{equation*}
\mathcal{J}_0=\begin{bmatrix}
	0         & 1  & 0         & 0  \\
	-f_{u0} & -c & -f_{v0} & 0  \\
	0         & 0  & 0         & 1  \\
	-g_{u0} & 0  & -g_{v0} & -c
\end{bmatrix},
\end{equation*}
where
\begin{align*}
&f_{u0} = 1-b-2u_0=-(1-b)\,,\quad  &f_{v0}&=(d-1)u_0-d\,,\\
&g_{u0} = 0\,,\quad  &g_{v0}&=a(1-u_0)-b\,d=(a-d)\,b\,,
\end{align*}
and the characteristic polynomial of $\mathcal{J}_0$ is
\begin{equation*}
x^2(c+x)^2+x(c+x)(f_{u0}+g_{v0})-f_{v0}\,g_{u0}=0\,.
\end{equation*}
This gives four distinct real eigenvalues:
\begin{equation*}
\begin{aligned}
&\lambda_{1,2}^{(0)} = \frac{1}{2}\Big(-c\pm\sqrt{c^2-4f_{u0}}\,\Big)\,,\quad f_{u0}<0~~(b<1 \text{ from } \cref{eq:trans})\,,\\
&\lambda_{3,4}^{(0)} = \frac{1}{2}\Big(-c\pm\sqrt{c^2-4g_{v0}}\,\Big)\,,\quad g_{v0}<0~~(a<d \text{ from } \cref{eq:trans})\,.\\
\end{aligned}
\end{equation*}
Thus, we have $\lambda_2^{(0)}<0<\lambda_1^{(0)}$ and $ \lambda_4^{(0)}<0<\lambda_3^{(0)}$, and steady state $E_0$ is a saddle point on the phase plane.

Repeating the analysis at $E_1$, we obtain the Jacobian matrix
\begin{equation*}
\mathcal{J}_1=\begin{bmatrix}
	0         & 1  & 0         & 0  \\
	-f_{u1} & -c & -f_{v1} & 0  \\
	0         & 0  & 0         & 1  \\
	-g_{u1} & 0  & -g_{v1} & -c
\end{bmatrix},
\end{equation*}
where
\begin{align*}
&f_{u1} = -b\,,~~&f_{v1}&=0\,,\\
&g_{u1} = -av_1 = b\,d-a\,,~~  &g_{v1}&=a-b\,d-2av_1=b\,d-a\,,
\end{align*}
and the eigenvalues of $\mathcal{J}_1$ are
\begin{equation*}
\begin{aligned}
&\lambda_{1,2}^{(1)} = \frac{1}{2}\Big(-c\pm\sqrt{c^2-4f_{u1}}\,\Big)\,,\quad f_{u1}=-b<0\,,\\
&\lambda_{3,4}^{(1)} = \frac{1}{2}\Big(-c\pm\sqrt{c^2-4g_{v1}}\,\Big)\,,\quad g_{v1}=b\,d-a <0~~(\text{from }\cref{eq:trans})\,.\\
\end{aligned}
\end{equation*}
Thus, we have $\lambda_2^{(1)}<0<\lambda_1^{(1)}$ and $ \lambda_4^{(1)}<0<\lambda_3^{(1)}$, and the steady state $E_1$ is also a saddle point on the phase plane. 

\begin{figure}[htbp]
\centering
\includegraphics[width =0.4\textwidth]{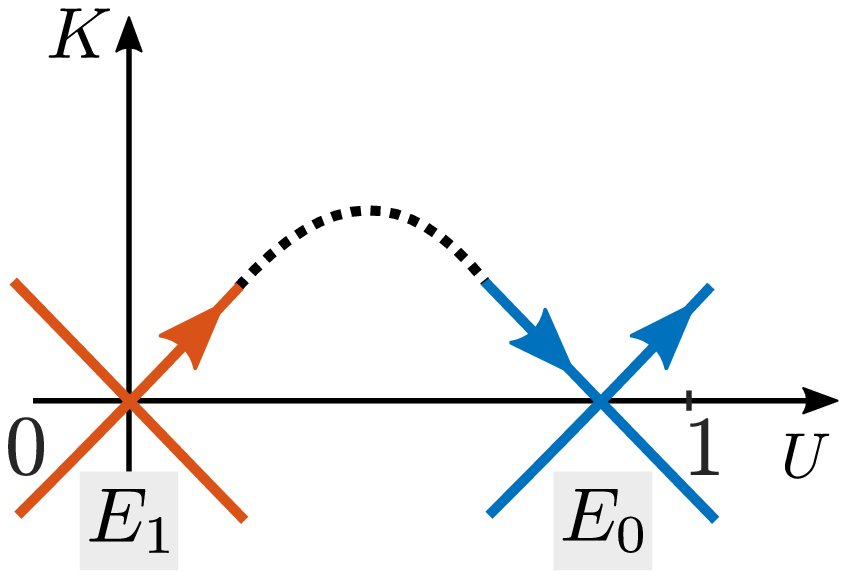}
\qquad\includegraphics[width =0.4\textwidth]{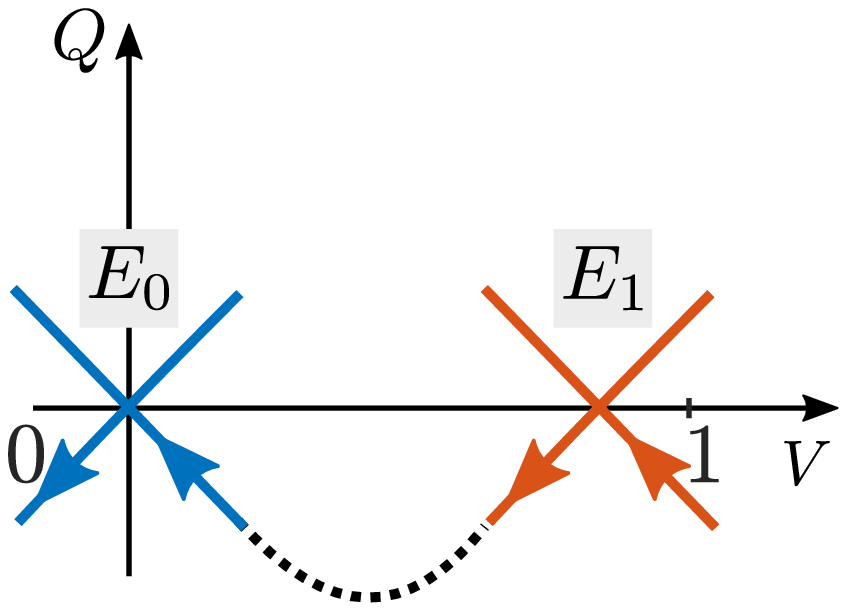}\caption{Phase plane analysis and sketch of solution trajectories\label{fig:phase}}
\end{figure}

Given that both $E_0$ and $E_1$ are saddle points, we sketch the phase plane trajectories and the steady states in \cref{fig:phase}. Notice that, we only consider the physically relevant trajectories within the space $0\le U\le1, K>0, 0\le V\le 1, Q<0$, that is the first quadrant stripe on the $U$-$K$ plane and the fourth quadrant strip on the $V$-$Q$ plane. Near the steady state $E_0$ (the blue trajectories), since $U$ is increasing and $K>0$, on the $U$-$K$ plane, the trajectory goes from left to right (in positive $U$-direction), while on the $V$-$Q$ plane, since $V$ is decreasing, the trajectory near $E_0$ goes from right to left (in negative $V$-direction). Similarly, we could determine the direction of trajectories near the saddle steady state $E_1$ (the orange ones). By continuity arguments, or by heuristic reasoning from the phase plane sketch of the trajectories, we claim that there is a trajectory that connects the steady states, which corresponds to the traveling wave front.

\subsection{Traveling wave speeds and shapes\label{sec:traveling}}
We are going to analyze the traveling wave profile and wave speed using the reduced 1-PDE model. We then compare the results with the numerical solutions of the 2-PDE model.
\subsubsection{Traveling wave solution for 1-PDE model}
\paragraph[For D=1]{For $D=1$}
We look for the traveling wave solution $p(x,t)=p(x-ct)=P(z)$ for the 1-PDE model \cref{eq:1-PDE}, which satisfies the ODE \cref{eq:travel}.
Let 
\begin{equation}
G(P)=\frac{dP}{dz}=-P'\,,
\label{eq:G}
\end{equation}
where the prime denote the derivative with respect to the $x$. We have picked the  coordinate direction $z=-x$, so that $P$ is increasing in $z$ and $G(P)\ge 0$. Then, $P''=(-G)_x=-G_x=-G_P P_z z_x=GG'$, and \cref{eq:travel} can be rewritten as an equation of variable $P$,
\begin{equation}
GG'-cG+h(P)=0\,,
\label{eq:G_system}
\end{equation}
with the boundary conditions
\begin{equation}
G(0)=G(1)=0\,.
\label{eq:G_system_BVP}
\end{equation}
We look for a wave speed $c>0$ that is consistent with the boundary value problem (BVP) \cref{eq:G_system,eq:G_system_BVP} using the linear shooting method. That is, for a given value $c$, we convert the BVP to an initial value problem (IVP) by using a linear approximation near $P=0$, and we identify the value $c$ such that the solution matches the boundary condition at the other end of the domain, $G(1;c)=0$. 
 
Suppose near $P=0$, we use linear approximate $G(P)\approx \lambda P$ ($\lambda>0$ since $G(P)\ge 0$). Substituting this approximation to \cref{eq:G_system}, we have 
\begin{equation*}
\lambda^2 P-c\lambda P+h(P)=0\,,
\end{equation*}
which gives
\begin{equation*}
\lambda_\pm = \frac{cP\pm\sqrt{(cP)^2-4Ph(P)}}{2P}\approx\frac{c\pm\sqrt{c^2-4h'(0)}}{2}~(\text{for small}~P)\,.
\end{equation*}
Since $h'(0)<0$ and $c>0$ (right-going wave), only the positive root, $\lambda_+>0$, is relevant. The second approximation is made for small $P$ near zero.

We now numerically integrate an IVP \cref{eq:G_system} subject to the initial condition
\begin{equation*}
G(P_\varepsilon) = \lambda_+\,P_\varepsilon\,, \quad \text{where}\quad P_\varepsilon\ll 1\,.
\end{equation*}
This gives $G(1;c)$ for any given $c$, and the solution for the original BVP problem, \cref{eq:G_system,eq:G_system_BVP}, corresponds to root of the nonlinear equation $G(1;c)=0$. Once the root is found (so as the $\lambda$), by the definition of $G(P)$ in \cref{eq:G}, we could integrate and obtain the traveling wave solution $P(-z)$.

\paragraph[For general D]{For general $D$}
For the general case \cref{eq:1-PDE-general}, it's a straightforward generalization of the $D=1$ case. Following the same idea, we have a BVP for $G_D(P)$
\begin{equation}
(D+(1-D)P)\,G_DG_D'-c\,G_D+h(P)=0\,,
\label{eq:G_system_D}
\end{equation}
subject to boundary condition 
\begin{equation}
G_D(0)=G_D(1)=0\,.
\label{eq:G_system_BVP_D}
\end{equation}
Consider a linear approximation at $P=0$, $G_D(P)\approx \lambda_D\,P$, and substitute it to \cref{eq:G_system_D}, we get one relevant positive coefficient (for small $P$)
\begin{equation*}
\lambda_D \approx\frac{c+\sqrt{c^2-4(D+(1-D)P)h'(0)}}{2(D+(1-D)P)}\,.
\end{equation*}
Thus, we transform the BVP, \cref{eq:G_system_D,eq:G_system_BVP_D}, into an IVP \cref{eq:G_system_D} with initial condition $G_D(P_\varepsilon)= \lambda_D\,P_\varepsilon,~ P_\varepsilon\ll 1$, and we solve the nonlinear equation $G_D(1;c)=0$ using iterative method.

\begin{figure}[htbp]
\centering
\includegraphics[width=0.45\textwidth]{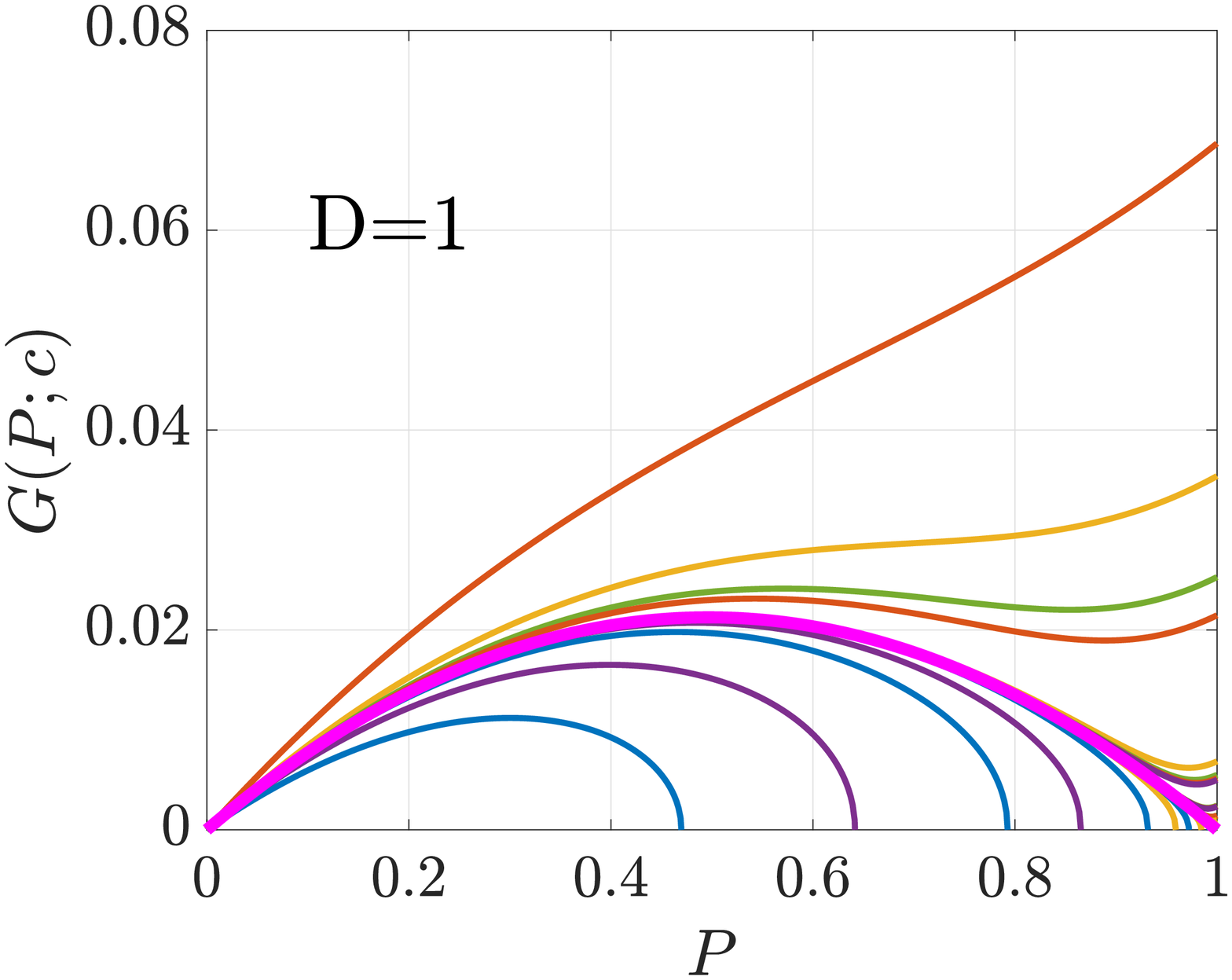}\includegraphics[width=0.45\textwidth]{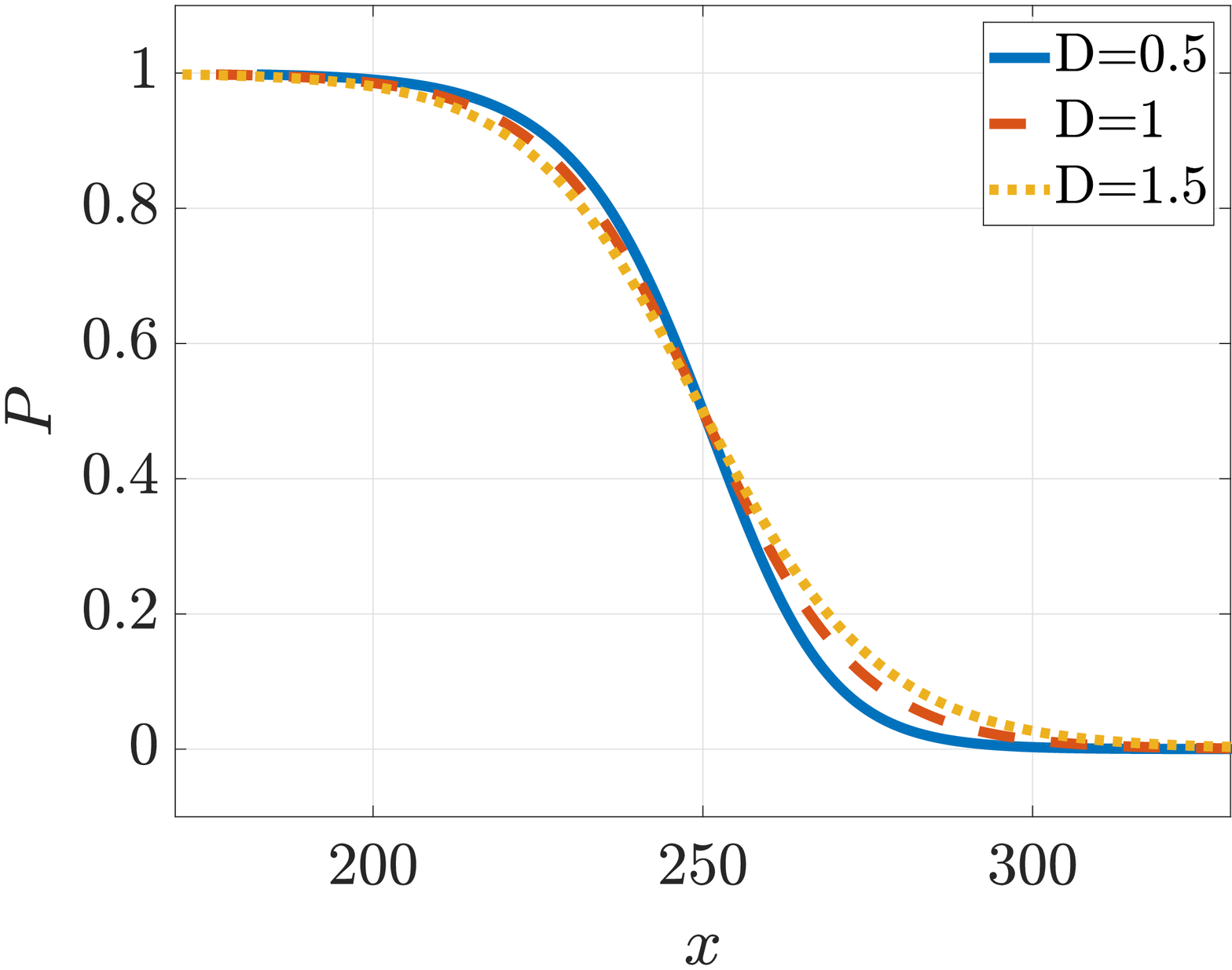}
\caption{\textbf{Left:} Iteration curves (on $c$) from shooting method, the final solution (corresponds to $c\approx0.046$) is marked in magenta, which passes (1,0) and satisfies the boundary value problem. \textbf{Right:} The traveling wave solution identified by the final result of the shooting method. \label{fig:shooting_curves}}
\end{figure}

In \cref{fig:shooting_curves} left, we plot the curve $G(P;c)$ at each iteration step when solving the root-finding problem at $D=1$. At the final iteration, the estimated wave velocity $c\approx0.046$, or equivalently $9.63$ m/day in the dimensional parameters. This gives a curve (in magenta) that satisfies $G(1;c)=0$ within the tolerance $10^{-6}$. On the right, we show the traveling wave solutions, $P(-z)=P(x)$, estimated by the shooting method for a range of $D$ values. As diffusion ratio $D$ increases, the wave shape becomes a bit slightly wider and flatter, and the front propagates faster. The corresponding estimated velocity is given in \cref{fig:wave_compare} right (1-PDE estimate curve). 

\subsubsection{Comparison with traveling wave solution for 2-PDE model}
\begin{figure}[htbp]
\centering
\includegraphics[width=0.45\textwidth]{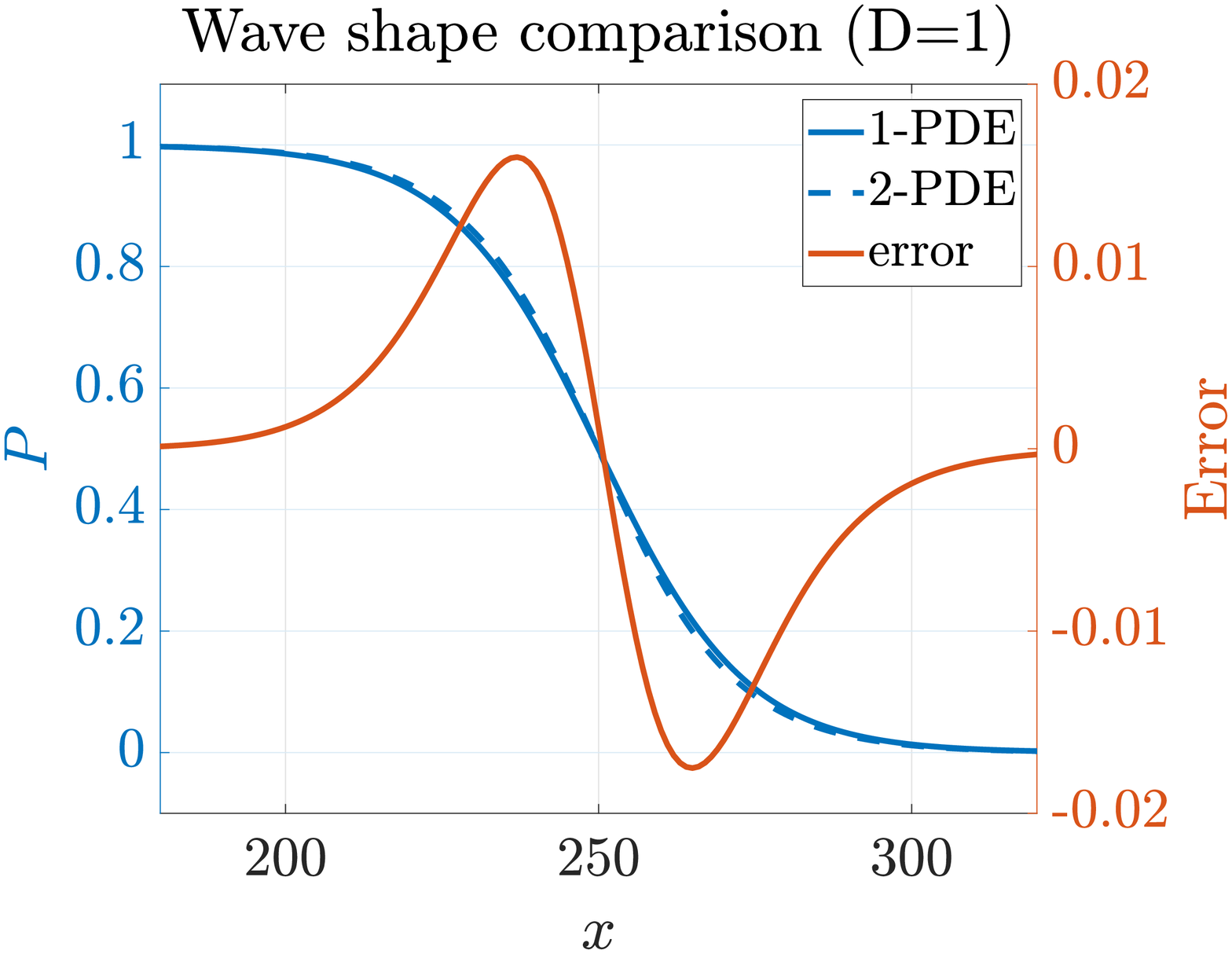}\quad\includegraphics[width=0.45\textwidth]{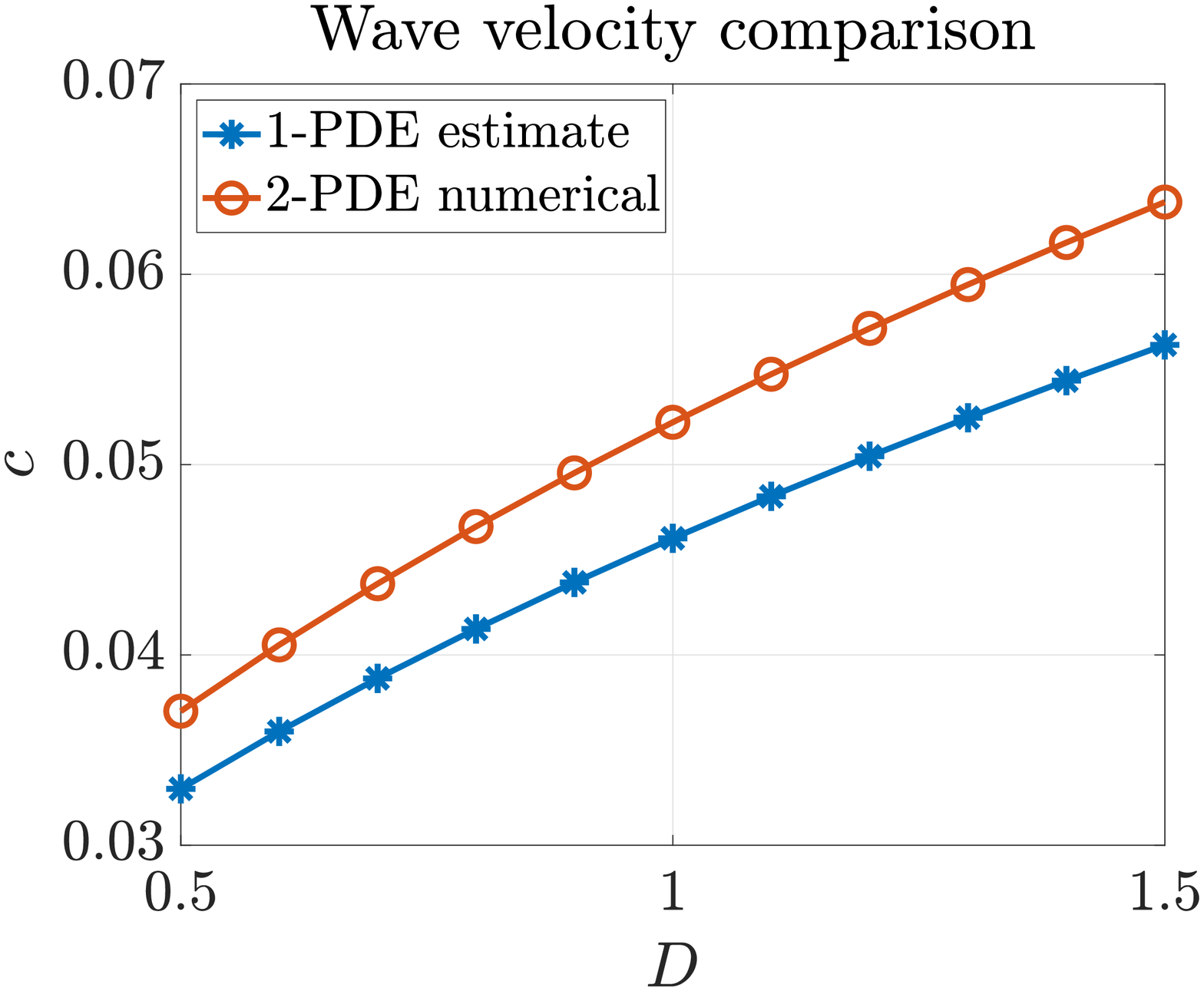}
\caption{Left: Comparison of traveling wave front for $D=1$. The left y-axis gives the infection level and right y-axis shows the discrepancy between two approaches. Right: Comparison of wave velocities. The 1-PDE estimates are consistently smaller than the 2-PDE reference. \label{fig:wave_compare}}
\end{figure}

We numerically integrate the 2-PDE model for a long time to obtain a reference for the traveling wave solution. At baseline $D=1$, we compare the shape of the infection fronts ($p=v/(u+v)$) with the 1-PDE result in \cref{fig:wave_compare} left. The solutions have been shifted in the x-coordinate to align in the center of the domain, and the error curve is plotted on the right y-axis. The traveling wave front obtained from two approaches are close, and the error $\|p_{1-PDE}-p_{2-PDE}\|_{\infty}\approx 0.016$. 

We numerically estimate the traveling wave velocity for the 2-PDE model by considering $ c(x,t) \approx -u_t(x,t)/u_x(x,t)$. We determine the velocity for the infection wave front when the median of $c(x,t)$ stabilizes in time and the wave front does not hit the computational domain. When $D=1$, the velocity $c\approx0.052$ (or $10.91$ m/day in the dimensional parameters). As seen from \cref{fig:wave_compare} right, the velocity from the 1-PDE model consistently underestimates the wave velocity for all the $D$ coefficients, and the relative error $(c_{2-PDE}-c_{1-PDE})/c_{2-PDE}\approx 12\%$.

\subsection{Practical consideration for successful invasion\label{sec:design}}
To establish a traveling wave of \W infection, we aim for an infection level above the critical bubble profile. The critical bubble is a threshold condition for wave initiation, however, it may not be an ideal release design if the faster establishment of the infection wave is desired. To inform a more practical scenario, we simulate releases of different target infection levels (as defined in \cref{sec:threshold2}) above the threshold. We search for an optimal level so that it balances the release time with the release amount needed for wave establishment. These simulations will be focused on the point-release strategy since it is a good approximation of a local release site. The insights gained from this simple setting may infer general principles that are applicable in other scenarios.

\textit{Minimal release time $T_1^*$.} We simulate the point-release scenario using a similar process described in \cref{sec:threshold2}. For each target infection level $\tilde{p}>p^\ast$: \textbf{step 1}, we release continuously (with boundary correction \cref{eq:fixBC}) at the release center for a period of time $T_1$; \textbf{step 2}, we stop releasing at the center and check if the traveling wave front could be established at time $T_2>T_1$. We vary the release time $T_1$ and iterate on step 1 \& step 2 to search for a minimal releasing time required. We use  a root-finding algorithm described in \cref{sec:appendix_minimal} to identify the $T_1^*$. The corresponding minimal release number for the point-release strategy is the total release number during step 1, that is $R(T_1^*)$, as defined in \cref{sec:release}.

\begin{figure}[ht]
\centering
\includegraphics[width=0.45\textwidth]{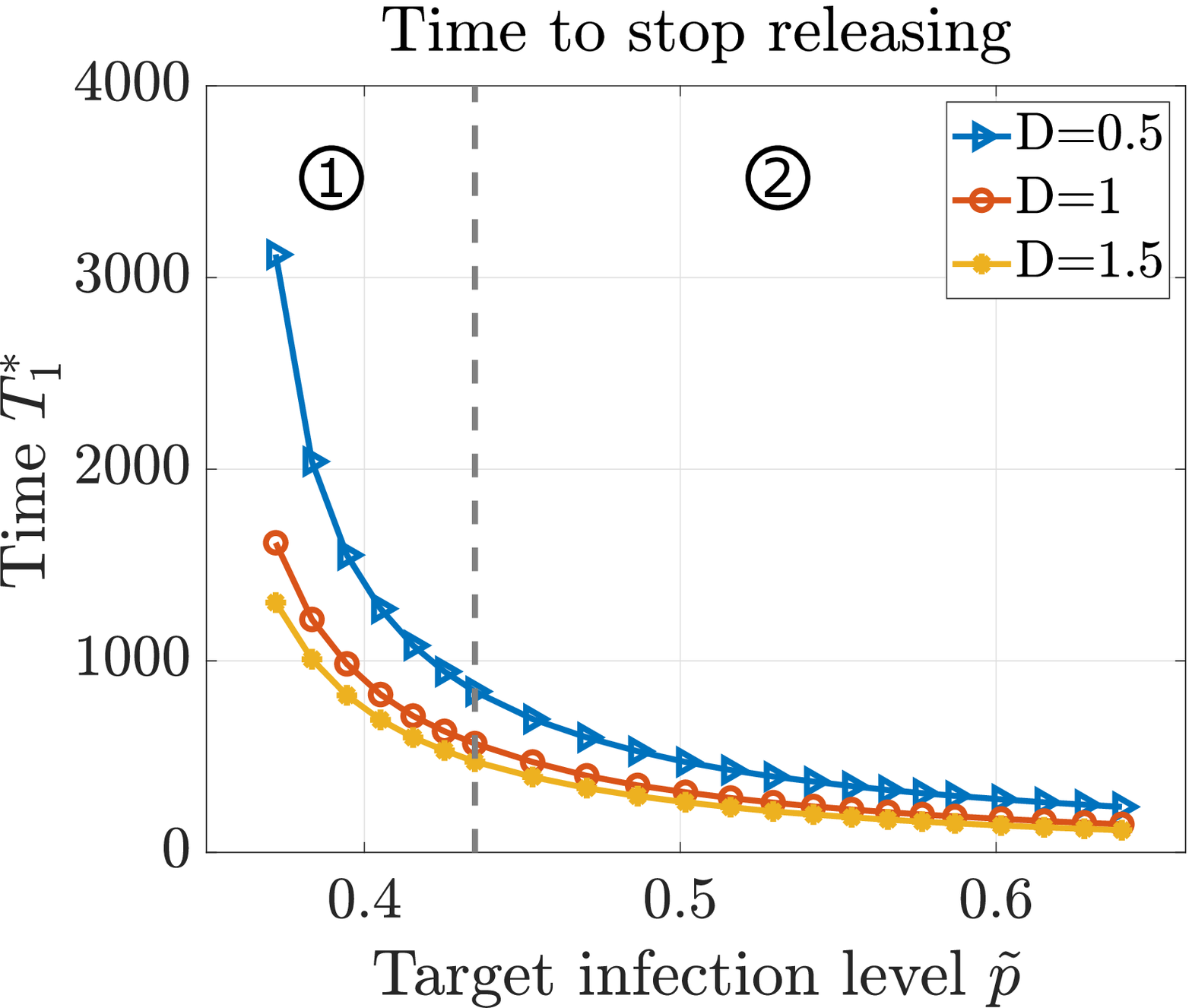}\qquad\includegraphics[width=0.45\textwidth]{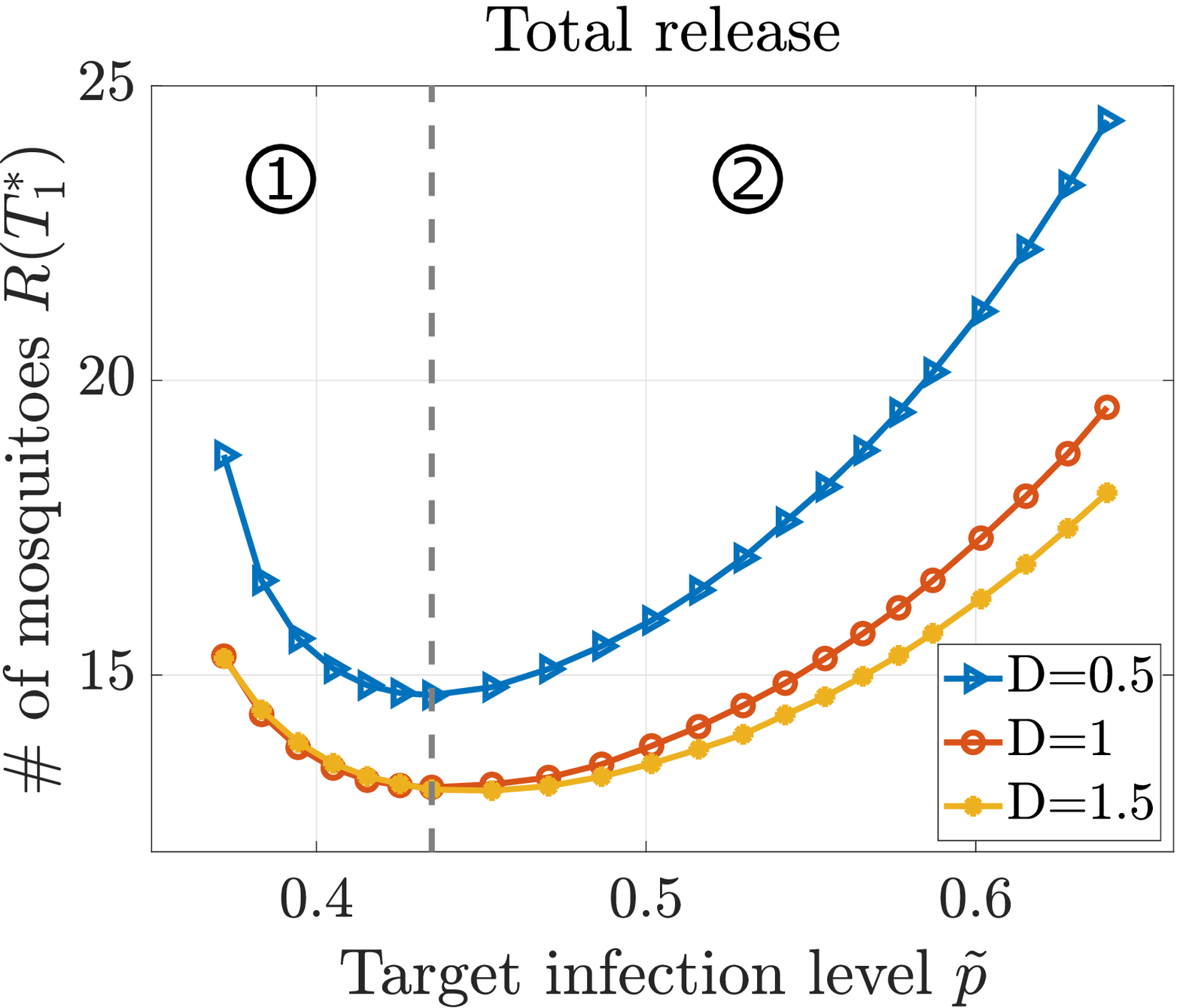}
\caption{Minimal release time (left) and total release number (right) using a point-release strategy. The optimal infection level to be maintained at the center is $\tilde{p}\approx 0.435$, which requires a shorter release time and smaller release size. \label{fig:release_travel}}
\end{figure}

\cref{fig:release_travel} left shows that increasing the target infection level $\tilde{p}$ results in a shorter minimal release time, but the reduction in time saturates and approaches a certain level for $\tilde{p}$ in the high-infection region \raisebox{.5pt}{\textcircled{\raisebox{-.9pt} {2}}}. For the total release curves (\cref{fig:release_travel} right), within the low-infection region \raisebox{.5pt}{\textcircled{\raisebox{-.9pt} {1}}}, although a larger $\tilde{p}$ requires a larger release initially at the release center, due to the benefit of the reduction in the release duration, the overall release number decreases. Meanwhile, in the high-infection region \raisebox{.5pt}{\textcircled{\raisebox{-.9pt} {2}}}, the release numbers bounce back. This is due to the penalty of the local carrying capacity in the model, and lots of the released infected mosquitoes die before they can diffuse into the nearby region to produce offspring. Thus, even the release amount increases for large $\tilde{p}$, it no longer improves the release time. Overall, to have a cost-effective release design, it's better to set a target infection level $\tilde{p}\approx 0.435$ ($1.89\times$ODE threshold level), so that it reduces the establishment time to a certain point but requires a relatively small number of infected mosquitoes.

We see a similar trend across different diffusion ratios $D$, and larger $D$ favors the establishment of the infection wave in terms of shorter minimal release time and smaller total release number. This is consistent with what we have observed for the establishment of the critical bubble (see \cref{sec:release} and \cref{fig:release_D}).

\section{Sensitivity Analysis\label{sec:SA}}
The model parameter values in \cref{tab:parameter_all} represent our baseline estimates, which inherent uncertainty from the biological measurements or depend on the choice of \W strains, mosquito species, local weather conditions, etc. We use sensitivity analysis to quantify the relative significance of the model parameters of interest (POIs) towards the output quantities of interest (QOIs). 

Following the framework in \cite{chitnis2008determining}, we define the normalized sensitivity index (SI) of a QOI, $q(p)$, with respect to the POI, $p$, as
\begin{equation*}
\mathcal{S}_p^q = \frac{p}{q}\times\frac{\partial q}{\partial p}\bigg|_{p=\hat{p}}\,,
\end{equation*}
at the baseline value $p=\hat{p}$. This dimensionless number predicts the impact of percentage change: if the parameter $p$ changes by $x\%$ around the baseline, then the quantity $q$ changes by $\mathcal{S}_p^q \times x\%$. To estimate the SI, we perturb the parameters (except $D_1$ and $D_2$) by $0.1\%$, and use centered difference to approximate the partial derivatives. For the diffusion coefficients $D_1$ and $D_2$, we have used $1\%$ perturbation to avoid any numerical instability, such as having a singular denominator in \cref{eq:HD}.

We also consider POIs that measure the fitness cost induced by the \W infection,
\begin{itemize}
\item[-] $r_{\mu} := (\mu_{fu}^{-1}-\mu_{fw}^{-1})/\mu_{fu}^{-1}$, which gives the fractional reduction in lifespan for the infected mosquitoes, and 
\item[-] $r_{\phi} := (\phi_{u}''-\phi_{w}'')/\phi_{u}''$, which gives the fractional reduction in the reproduction rate among the infected mosquitoes.
\end{itemize}
We present the SI results in \cref{tab:SA} for both the original 2-PDE model and the reduced 1-PDE model, and the reduced 1-PDE model preserves the order of significance and closely approximates the index values of the 2-PDE ones.

\begin{table}[htbp]
\centering
\caption{Normalized sensitivity indices for QOIs (top row) with respect to POIs (left column) for the 2-PDE model and reduced 1-PDE model.  \label{tab:SA}}
\begin{tabular}{c SS SS SS}
\toprule
& \multicolumn{2}{c}{PDE threshold} &\multicolumn{2}{c}{Bubble area} & \multicolumn{2}{c}{Wave speed}\\
\cmidrule(lr){2-3} \cmidrule(lr){4-5}\cmidrule(lr){6-7} 
& $\text{1-PDE}$ & $\text{2-PDE}$  & $\text{1-PDE}$ & $\text{2-PDE}$  & $\text{1-PDE}$ & $\text{2-PDE}$\\
\midrule
$v_w$ & -4.54 &-4.43 & -3.20 & -3.18&  5.09&5.19\\
\midrule
$\phi_u''$ & 3.40 & 3.28 & 2.27 &2.26 & -2.77& -2.66\\ 
\addlinespace
$\phi_w''$ &-3.40& -3.29 & -1.75 &-1.75&  2.27&2.16\\  
\addlinespace
$r_{\phi}$ & 0.79 & 0.76 & 0.40& 0.40&-0.53 &-0.49\\
\midrule
$\mu_{fu}'$ & -2.62 & -2.54&  -1.48&-1.09&2.09&1.85\\  
\addlinespace
$\mu_{fw}'$ & 2.62 & 2.54& 0.96&0.58& -1.59& -1.36\\ 
\addlinespace
$r_{\mu}$ & 0.18 & 0.18 &0.07 & 0.04 &-0.11 &-0.09\\
\midrule
$D_1$  & 0.03 & 0.04 & -0.34&-0.38& -0.49&-0.50\\
\addlinespace
$D_2$ & -0.03 &  -0.04 & 0.35&0.38& 0.49&0.50\\
\midrule
$K_f$ &0 & 0& 0&0& 0&0\\
\bottomrule
\end{tabular}
\end{table}

\subsection{Impact of imperfect maternal transmission\label{sec:imperfect}}
The maternal transmission rate, $v_w$, measures the fraction of infection among the offspring reproduced by the infected females, and it has been a significant parameter that impacts the threshold condition and invasion process in the spatially homogeneous setting \cite{qu2019generating,qu2018modeling}. For simplicity, we have based our previous discussions on the perfect maternal transmission rate $v_w=1$. To study the impact of the imperfect case, when $v_w<1$, we derive the corresponding threshold and traveling wave conclusions, which is a straightforward extension of the previous analysis. The results are summarized in \cref{sec:appendix}.

As in the ODE setting, the maternal transmission rate, $v_w$, is still the most sensitive parameter across all the QOIs for the spatial models. The magnitude of the SI for the PDE threshold is comparable to the ODE setting ($-4.36$ in \cite[Table 6.1]{qu2018modeling}).

\subsection{Sensitivity analysis on other model parameters}
From \cref{tab:SA}, the magnitudes of the SIs for the reproduction rates ($\phi_*''$) is greater than the ones for the death rates ($\mu_*'$). This suggests that the reproduction of offspring is more important than the lifespan of the mosquitoes when it comes to the invasion process, including determining the threshold infection level needed at the release center and predicting the propagation speed for the infection wave.

This trend could be better observed by considering the relative reductions in the reproduction and lifespan due to the \W infection. From the SI table, we have
$SI_{r\phi}^*/SI_{r\mu}^*>4$ for all the QOIs. This indicates that the impact of the reduction in reproduction rate, as measured by the magnitudes of the SI, is more than $4\times$ greater than the reduction in lifespan. Specifically, for every 1\% of reduction in the reproduction rate, it would raise the threshold by 0.76\%, while for lifespan, the increase is 0.18\%; Similarly, 1\% of reduction in reproduction rate will slow the invasion front by 0.49\%, while the 1\% decrease in lifespan will slow the front by 0.09\%. We could see a similar comparison in the ODE setting \cite[Table 6.1]{qu2018modeling}, however, the difference is much smaller (less than $2\times$). In another word, the spatial models illustrate the important role that reproduction rates play in the invasion process.

The sensitivity analysis results also suggest that a smaller diffusion coefficient, or the decrease in the flying activities, among the infected mosquitoes may increase the invasion threshold and make it harder to spread out the infection. However, the relative impact is not as significant as the other parameters discussed before.

Lastly, all the QOIs are not sensitive to the change in carrying capacity, $K_f$. This is because the invasion dynamics are only determined by the competition between the infected and uninfected mosquitoes. Our model formulation has assumed that the two types of mosquitoes are equally impacted by the $K_f$, thus changing $K_f$ won't affect the density of infection. 

\section{Discussions and Conclusions}
We created and analyzed spatial models for \W invasion dynamic in the field. The 2-PDE model is based on previous ODE models, where there exists a critical threshold infection level for the infected mosquitoes to persist in the population. We derived the spatial models to better describe the heterogeneity in field releases that comes from the local introduction of infection and mosquito random flights. This extension leads to nontrivial changes in its biological dynamics and provides key insights for the field trial design.

We proposed a 2-PDE reaction-diffusion model for the infected and uninfected mosquitoes. This system was further simplified into a 1-PDE model for the infection density in the mosquitoes to better understand the dynamics of the complex 2-PDE system. We derived analytical results using the more manageable 1-PDE model and compared them to the numerical results of the more accurate 2-PDE model. 

We first identified the threshold condition for establishing \W invasion wave, given a local release of infection. The obtained threshold condition is realized as a bubble-shaped spatial distribution of infection, referred to as a critical bubble. Our numerical results suggest that the critical bubble, which balances the reproduction and diffusion dynamics, is an optimal spatial distribution of the infection to sustain the infection, compared to other spatial configurations.

Moreover, the infection level at the release center of the balanced critical bubble (PDE threshold) is higher than the ODE threshold ($p_{PDE}^*\approx 0.35$ vs. $p_{ODE}^*\approx 0.23$). This illustrates the impact of the non-homogeneous mixing between the infection groups and confirms the necessity of using the more realistic spatial models for predicting the \W field releases.

When above the threshold condition, the proposed models give rise to the traveling wave solutions. We analyzed the wave speed and the shape of the wave front using both the 1-PDE and 2-PDE models. At the baseline, the wave speed is $c\approx0.046$, or $9.63$ m/day in the dimensional parameters. 

Our conclusions and calculations are based on the baseline parameters, which are our best-guess estimates but naturally involve bias and uncertainty. Our sensitivity analysis showed that the maternal transmission rate is the most important parameter during the invasion process, including the threshold condition and traveling wave speed. The results also uncover that the reproduction rates have a much larger impact than the mosquito lifespan for the invasion. This may inform the choice of different \W strains with different levels of fitness costs on the infected mosquitoes.

This study is our preliminary attempt to explore how the spatial dynamic may affect the prediction of \W field releases, which offers important insights that would be otherwise neglected under the ODE setting. However, there are lots of assumptions that we have made to be mathematically tractable. One major assumption is that we only tracked the adult mosquitoes since our model is based on a 2-ODE model that has been derived from a 9-ODE model through a model reduction process. This leads to the caveat that the current models may not be suitable to predict field trials that break the natural balance among different life stages or the sexual ratio of mosquitoes. In such a case, it may be worthwhile to derive from the full 9-ODE model to include aquatic and male compartments for simulation purposes.

Furthermore, before applying this model to guide field releases of infected mosquitoes, the model must be extended to two spatial dimensions, where the infected mosquitoes are released in a symmetrical bubble and the infection wave propagates in a circular motion. Our future work will be to determine how the threshold condition adapts accordingly in this case and see if the \W infection could be sustained at the front of the infection wave.

\appendix
\section{Capturing the threshold for 2-PDE model - Step 3\label{sec:appendix_step3}} The critical bubble distribution of infected mosquitoes is an unstable equilibrium solution of the PDE model. Due to the instability and the stiffness of the system near this state, it is a challenging numerical problem to identify $p^\ast$ in step 3 of the algorithm described in \cref{sec:design}. We design the following root-finding problem to numerically approximate the threshold condition with high accuracy.

The key to constructing a robust objective function for iteration is to characterize the distinct dynamics when the infection level is above or below the threshold:
\begin{itemize}[leftmargin = \parindent]
\item[-] When $\tilde{p}$ is slightly above the threshold $p^*$, let $T_1\rightarrow\infty$ in step 1, and the infection forms an unstable front (close to the critical bubble, but not converging to it) for a while. Eventually, the unstable infection curve grows and approaches the upper stable steady state, which creates a boundary layer at $x=0$ due to the boundary correction \cref{eq:fixBC}.
\item[-] When $\tilde{p}$ is slightly below the threshold $p^\ast$, let $T_1\rightarrow\infty$, the infection converges and forms a stable bubble. Once the boundary condition is relaxed in step 2, the infection collapses (as $T_2\rightarrow\infty$).
\end{itemize} 
Employing these two observations, we design the following root-finding problem, which is solved using the bisection method:
\begin{equation*}
\mathcal{J}(\tilde{p})=1-2\times\underbrace{\left\{p(0,T_2;\tilde{p})<\tilde{p}\right\}}_{\text{Condition I}}\times\underbrace{\left\{\frac{\|(p(\cdot,T_1;\tilde{p})-p(\cdot,T_1-\dt;\tilde{p})\|_{l_2}}{\|(p(\cdot,T_1-\dt;\tilde{p})\|_{l_2}}<10^{-4}\right\}}_{\text{Condition II}} =0\,,
\end{equation*}
where $\Delta t$ is the step size for temporal discretizations, and $T_1$ and $T_2$ are taken to be sufficiently large. The brackets around the conditions gives 1 or 0 value, when the condition is true or false, respectively. 

Condition I in the objective function $\mathcal{J}(\tilde{p})$ checks the infection level at the release center, and condition II checks the relative convergence of the infection front. For $\tilde{p}>p^*$, condition I may fail if $\tilde{p}\gg p^*$, and condition II may fail if $\tilde{p} \gtrapprox  p^*$, thus $\mathcal{J}(\tilde{p}) = 1$; For $\tilde{p}<p^*$, condition I holds, and condition II holds for large $T_1$, thus $\mathcal{J}(\tilde{p}) = -1$. Although there is no exact root for $\mathcal{J}(\tilde{p})=0$, by applying the bisection method, we obtain an estimate for the threshold $p^\ast$ within an error tolerance $<10^{-5}$.

\section{Identifying minimal release time for sustained infection\label{sec:appendix_minimal}} 
As described in \cref{sec:design}, to identify the minimal release time, $T_1^*$, for the point-release process, we iterate on the duration of step 1 (release time $T_1$) such that the infection could be sustained and established in step 2 (final time $T_2$). The iteration can be summarized by the following root-finding problem:
\begin{align*}
\mathcal{K}(T_1;\tilde{p})=2\times \left\{p(0,T_2;T_1, \tilde{p})>\tilde{p}\right\}-1=0\,.
\end{align*}
Here, $\tilde{p}$ is the target infection level at the release center, and brackets operation returns 1 or 0 values depending on if the condition inside is satisfied or not. If $T_1 > T_1^*$, then the traveling wave will be established and the infection rate at release center will be greater in step 2 ($\mathcal{K}(T_1;\tilde{p})=1$); If $T_1 < T_1^*$, the infection will collapse in step 2, and 
$\mathcal{K}(T_1;\tilde{p})=-1$. In our numerical simulations, we use $T_2=T_1+10^6$. Similar to the problem defined in \cref{sec:appendix_step3}, although there is no exact root for $\mathcal{K}(T_1;\tilde{p})=0$, given a fine enough time discretization $\Delta t$ in step 1, we could find an estimate for $T_1^*$ within the tolerance $< \Delta t$.

\section{Conclusions for imperfect maternal transmission\label{sec:appendix}}
For a general maternal transmission rate $v_w$, the 2-PDE model is written as
\begin{align*}
\frac{\partial F^{u}}{\partial t} &= b_f \phi_u''\,\frac{F^u}{F^u+\frac{\mu_{fw}'}{\mu_{fu}'}F^w} \left(1-\frac{F^u+F^w}{K_f}\right) F^u\\
&\qquad \qquad\quad+v_u b_f \phi_w'' \left(1-\frac{F^u+F^w}{K_f}\right) F^w  -\mu_{fu}' F^u+\nabla\cdot (D_1 \nabla  F^u)\,,\\
\frac{\partial  F^{w}}{\partial t} &= v_w b_f \phi_w'' \left(1-\frac{F^u+F^w}{K_f}\right) F^w-\mu_{fw}' F^w+\nabla\cdot (D_2 \nabla  F^w)\,,
\end{align*}

and the corresponding nondimensionalized system \cref{eq:2PDE} is modified as 
\begin{equation}\label{eq:2PDEi}
\begin{aligned}
u_t &= \frac{u}{u+d\,v} (1-u-v)u +(1-m) a (1-u-v) v -b\,u+ u_{xx}\,,\\
v_t &= m a (1-u-v) v-b\,d\,v+Dv_{xx}\,,
\end{aligned}
\end{equation}
where we have used the notation $m=v_w$ to avoid the confusion with the state variable $v$. The reduced 1-PDE model could be obtained by modifying the transformation \cref{eq:trans2} as $ u+v = 1-\frac{bd}{m a}+\varepsilon\,, \frac{v}{u+v} = p\,,$ and the 1-PDE equation \cref{eq:1-PDE} becomes
\begin{equation}\label{eq:1-PDEi}
\begin{aligned}
&p_t = h_m(p) +p_{xx}\,,\\
&h_m(p)=\,\frac{bp\left(a m (1-p)^2+a d^2 (m-1) p^2-d (p-1) (a (2 m-1) p+p-1)\right)}{am(1+(d-1)p)}\,.
\end{aligned}
\end{equation}

\paragraph*{Conclusions for threshold conditions} For the 1-PDE model \cref{eq:1-PDEi}, the threshold condition $p^*_m$ is the root for the nonlinear equation
\begin{align*}
&H_m(p)=\int_0^p h_m(y) dy = -b\big((d-1) p \big(2 (d-1)^2 p^2 \left(d (a d-a+1)-a (d-1)^2 m\right)-\\
& 3 (d-1) p \left(a (d-1)^2 m+2d^2-d\right)+6 d^3\big)-6 d^3 \log ((d-1) p+1\big)/(6 a (d-1)^4 m)=0\,,
\end{align*}
and the critical bubble satisfies the IVP $ p'(x)=-\big(-2H_m(p)\big)^{1/2}$, $p(0)=p_m^*$.

For the 2-PDE threshold, the numerical algorithms described in the main text  (\cref{sec:threshold2}) can be applied to \cref{eq:2PDEi} without modifications.
\paragraph*{Conclusions for traveling wave}  The methods and algorithms discussed in \cref{sec:traveling} can be applied to both the 1-PDE and 2-PDE models here without changes.
\section*{Acknowledgments} 
Z. Qu and T. Wu were supported by the University of Texas at San Antonio New Faculty Startup Funds.  This research was partially supported by the NSF award 1563531. The content is solely the authors' responsibility and does not necessarily represent the official views of the National Science Foundation.

\bibliographystyle{siamplain}
\bibliography{WolbachiaPDE}


\end{document}